\documentclass[rnote]{aa} % for the research notes
\usepackage{graphicx}
\usepackage{txfonts}
\usepackage{natbib}
\bibpunct{(}{)}{;}{a}{}{,} % to follow the A&A style
\begin{document}
\title{Discovery of ``isolated'' comoving T Tauri stars in Cepheus\thanks{Based 
on observations collected at the \textit{Observatoire de Haute Provence} (OHP, France) and the 
\textit{Catania Astrophysical Observatory} (OAC, Italy).}
\fnmsep\thanks{Figure \ref{Fig:PTTSCCF} is only available in electronic form at {\tt http://www.aanda.org}}}

\author{P. Guillout \inst{1},
	A. Frasca\inst{2},
	A. Klutsch \inst{3,1}, 
	E. Marilli\inst{2},
	\and
	D. Montes\inst{3} 
          }

\offprints{P. Guillout\\ \email{patrick.guillout@astro.unistra.fr}}

\institute{
Observatoire Astronomique, Universit\'e de Strasbourg \& CNRS, UMR 7550, 11 rue de l'Universit\'e, 67~000 Strasbourg, France
\and
INAF - Osservatorio Astrofisico di Catania, via S. Sofia, 78, 95123 Catania, Italy
\and
Departamento de Astrof\'{\i}sica y Ciencias de la Atm\'osfera, Universidad Complutense de Madrid, E-28040 Madrid, Spain
}

\date{Received 15 April 2010 / Accepted 19 July 2010}
 
\abstract 
  % context heading (optional), leave it empty if necessary 
{During the course of a large spectroscopic survey of X-ray active late-type stars in the solar neighbourhood, we discovered four lithium-rich stars packed within just a few degrees on the sky. Although located in a sky area rich in CO molecular regions and dark clouds, the Cepheus-Cassiopeia complex, these very young stars are projected several degrees away from clouds in front of an area void of interstellar matter. As such, they are very good ``isolated'' T Tauri star candidates.}
  % aims heading (mandatory)
{We present optical observations of these stars conducted with 1--2 meter class telescopes. We acquired high-resolution optical spectra as well as photometric data allowing us to investigate in detail their nature and physical parameters with the aim of testing the ``runaway'' and ``in-situ'' formation scenarios. Their kinematical properties are also analyzed to investigate their possible connection to already known stellar kinematic groups.}
  % methods heading (mandatory)
{We use the cross-correlation technique and other tools developed by us to derive accurate radial and rotational velocities and perform an automatic spectral classification. The spectral subtraction technique is used to infer chromospheric activity level in the H$\alpha$ line core and clean the spectra of photospheric lines before measuring the equivalent width of the lithium absorption line.}
  % results heading (mandatory)
{Both physical (lithium content, chromospheric, and coronal activities) and kinematical indicators show that all stars are very young, with ages probably in the range 10--30\,Myr. In particular, the spectral energy distribution of TYC\,4496-780-1 displays a strong near- and far-infrared excess, typical of T Tauri stars still surrounded by an accretion disc. They also share the same Galactic motion, proving that they form a homogeneous moving group of stars with the same origin.}
 % conclusions heading (optional), leave it empty if necessary 
{The most plausible explanation of how these ``isolated'' T Tauri stars formed is the ``in-situ'' model, although accurate distances are needed to clarify their connection with the Cepheus-Cassiopeia complex. The discovery of this loose association of ``isolated'' T Tauri stars can help to shed light on atypical formation processes of stars and planets in low-mass clouds.}

\titlerunning{Discovery of ``isolated'' comoving T Tauri stars in Cepheus}
      \authorrunning{P. Guillout et al.}

\maketitle

%===================================================================
\section{Introduction}
\label{Sec:intro}

Although most studies of very young low-mass stars have concentrated on star-forming regions (SFRs) or young clusters, it has become increasingly evident that a considerable number of very young stars unrelated to prominent SFRs is  present in the solar neighbourhood. Their existence far from interstellar material is difficult to reconcile with the  standard picture of star formation. Different scenarios, such as ``runaway'' T\,Tauri stars \citep{1995A&A...304L...9S, 1998A&A...339...95S} or formation in small turbulent clouds \citep{1996ApJ...468..306F}, have been proposed as possible explanations. Most of the comoving associations of nearly coeval stars (i.e., moving groups (MG)) discovered so far are located in the southern hemisphere and may be tied to the Scorpius-Centaurus association (see, e.g., \citealt{2004ARA&A..42..685Z} and \citealt{2008hsf2.book..757T}).

%-----------------------  Table with literature data  ----------------------------------
\begin{table*}[t]
%\vspace*{-1.0cm}
\caption{Main data of our sources from the literature and dates of observations.}
\begin{tabular}{llccllccc}
\hline
\hline
Name & RasTyc source & $\alpha$ (2000) &  $\delta$ (2000) &  $V_{\rm T}$ & $(B -V)_{\rm T}$ & PSPC  & Li obs. &  H$\alpha$ obs.\\ 
& & h m s & $\degr ~\arcmin ~\arcsec$ & & & (ct/s) & &\\ 
\hline
BD+78\,853 & RasTyc0000+7940 &  00 00 38.4  & +79 40  36.9 & 10.35 & 0.79 & 0.10 & 2001/11/10 & 2001/11/20 \\
TYC 4496-780-1 & RasTyc0013+7702  & 00 13 40.5 & +77 02 10.9 & 9.83 & 0.70 & 0.10 & 2001/11/12 & 2001/11/21 \\
TYC 4500-1478-1 & RasTyc0038+7903 & 00 38 06.0 & +79 03 28.5 & 10.37 & 1.04 & 0.11 & 2003/12/14 & 2009/10/06\\
BD+78\,19 & RasTyc0039+7905 & 00 39 40.1 & +79 05 30.8 & 9.67 & 0.77 & 0.05 & 2002/11/02 & 2002/11/05 \\
\hline
\end{tabular}
\label{Tab:StarPos}
%\begin{list}{}{}
%\item$^{\mathrm{a}}$ ROSAT Position Sensitive Proportional Counter.
%\end{list}
\end{table*}
%
%-----------------------  Table with Photometric data  ----------------------------------
\begin{table*}
\caption{Johnson-Cousins and Str\"omgren photometric data.}
\begin{tabular}{lcccccccc}
\hline
\hline
Name & $V$ & $R_{\rm C}$ & $I_{\rm C}$ & $B-V$ & $U-B$ & $b-y$ & $m_1$ & $c_1$\\
\hline
BD+78\,853       & 10.29 (0.05)           & 9.77$^{\rm a}$ & 9.42$^{\rm b}$ & 0.62 (0.02) & 0.16 (0.02) & 0.462 & 0.164 & 0.266 \\
TYC 4496-780-1  & 9.78 (0.05)            & 9.22$^{\rm a}$ & 8.858$^{\rm c}$ & 0.64 (0.03) & 0.15 (0.02) & 0.438 & 0.211 & 0.269 \\
TYC 4500-1478-1 & 10.47$^{\rm a}$ (0.03) & 9.96$^{\rm d}$ (0.03) & 9.60$^{\rm d}$ (0.02) & 0.87 (0.03) & 0.40 (0.02) & 0.572 & 0.243 & 0.400 \\
BD+78\,19        & 9.57 (0.03)            & 9.11 (0.03) & 8.89 (0.02) & 0.72 (0.03) & 0.09 (0.02) & 0.457 & 0.218 & 0.260\\
\hline
\end{tabular}
\label{Tab:StarPhot}
\begin{list}{}{}		         	                	                 
\item[$^{\mathrm{a}}$] \cite{2004AAS...205.4815Z} \hspace{0.5cm} $^{\mathrm{b}}$ \cite{2003AJ....125..984M} \hspace{.5cm} $^{\mathrm{c}}$ \cite{2006PASP..118.1666D} \hspace{.5cm} $^{\mathrm{d}}$ Corrected from the contribution of the nearby companion.
\end{list}
\end{table*}

Single active stars in the field, selected on the basis of their high coronal emission, are mostly young stars with an age of a few 100\,Myr, i.e. in the zero-age main sequence (ZAMS), or even younger: post-T Tauri stars (PTTS) or T Tauri stars (TTS). \citet{2009A&A...504..829G} conducted a large ground-based observing program designed to perform the physical characterization of $\approx$~1000 {\it RasTyc} stellar X-ray sources \citep{1999A&A...351.1003G} by analysing their high-resolution optical spectra. Although the sky density of the youngest stars (identified thanks to their very high-lithium content) is more or less uniform \citep{2008PhDT.........7V}, we discovered an unusual {\it group} of four lithium-rich stars towards the Cepheus-Cassiopeia (Cep-Cas) complex \citep[see, e.g.,][and references therein]{2008hsf1.book..136K}. However, their space distribution in a wide region devoid of dense CO clouds and their stellar properties make the relation between these stars and the Cep-Cas complex uncertain. Young stars in the sky region around the Cep-Cas complex are indeed not necessarily associated with these SFRs. A good example is represented by the nearby young visual binary HIP\,115147 (V368 Cep) and its reported comoving companion \citep{2007ApJ...668L.155M}, which are also projected towards the same region of the sky. HIP\,115147 is currently classified as a very young ``naked'' PTTS \citep{1990BAAS...22.1254N,1991IBVS.3623....1C,1993IzKry..88...29C}, 20 to 50 Myr old, located at 20\,pc whose origin remains disputed although it is certainly unrelated to the Cep-Cas complex \citep{2001A&A...379..976M, 2010A&A...514A..97L}.\\

In the present paper, we concentrate on the four stars that we previously discovered and study their properties, evolutionary status, and kinematics. The observations and data reduction are outlined in Sect.~\ref{Sec:Data}. The determination of their stellar parameters as well as the level of chromospheric activity, lithium content, and age are briefly discussed in Sect.~\ref{Sec:Res}. We discuss the properties of each star under scrutiny in Sect.~\ref{Sec:Disc}. We conclude and summarize the results from the ongoing study of these sources in Sect.~\ref{Sec:Conc}.
\section{Observations and data reduction}
\label{Sec:Data}

Spectroscopic observations were conducted in 2001, 2002, and 2003 (see Table~\ref{Tab:StarPos}) at the OHP 1.52-m telescope using the \textit{Aurelie} spectrograph to acquire spectra in both the H$\alpha$ ($\lambda\lambda\approx$ $6500-6620\,\AA$) and lithium ($\lambda\lambda\approx$ $6650-6775\,\AA$) spectral regions at a resolution of $R\approx$~38\,000. An echelle spectrum at $R \approx 21\,000$ of TYC 4500-1478-1 was also taken with the FRESCO spectrograph of the 0.91-m telescope of the OAC in 2009. For details about the reduction of \textit{Aurelie} and FRESCO spectra, the reader is referred to \cite{2009A&A...504..829G}. 

In 2008, complementary photometric observations were performed in the standard $UBV$ system and Str\"omgren {\it uvby} filters with the 91-cm telescope of the OAC and a photon-counting refrigerated photometer equipped with an EMI 9893QA/350 photomultiplier, cooled to $-15^\circ$C. For details about the reduction of photometric data, the reader is referred to \cite{2006A&A...454..301F}. CCD images in Johnson-Cousins $V$, $R_{\rm C}$, and $I_{\rm C}$ filters were also acquired for two stars using the OAC focal-reducer imaging camera. The reduction of these data was performed following standard methods and the $VR_{\rm C}I_{\rm C}$ magnitudes were extracted from the corrected images by means of aperture photometry. Photometric data are summarized in Table~\ref{Tab:StarPhot}.

\begin{table*}[t]
\caption{Astrophysical parameters from spectroscopy and Str\"omgren photometry, lithium abundance, and chromospheric/coronal luminosities.}
\begin{tabular}{lcccccccccc}
\hline
\hline
Name & $T_{\rm eff} ^{\rm 1}$ & $\log g^{\rm 1}$ & [Fe/H]$^{\rm 1}$ & $T_{\rm eff} ^{\rm 2}$ & $R^{\rm 2}$ & $\delta m_0^{\rm 2}$ & $E(b-y)^{\rm 2}$ & $W_{\rm Li}$ / $\log N{\rm(Li)}$ & $W_{\rm H\alpha}^{em}$ /  $\log L_{\rm H\alpha}$ & $\log L_X$\\
         & (K) & & & (K) & (R$_{\sun}$) & & (mag) & ($\AA$) / &   ($\AA$) / (erg\,s$^{-1}$) & (erg\,s$^{-1}$)\\
\hline
BD+78\,853      & 5738 & 4.19 & $-0.03$ & 5750 & 0.98 & 0.032 & 0.068 & 0.18 / 3.05 & 0.31 / 29.5 & 30.6 \\
TYC 4496-780-1  & 5844 & 4.17 & $-0.04$ & 5650 & 0.99 & 0.012 & 0.024 & 0.18 / 3.14 & 3.90 / 30.6 & 30.4 \\
TYC 4500-1478-1 & 5160 & 4.30 & $-0.06$ & 5330 & 0.88 & 0.052 & 0.099 & 0.30 / 3.09 & 0.97 / 29.6 & 30.4 \\
BD+78\,19       & 5444 & 4.14 & $-0.11$ & 5630 & 0.96 & 0.003 & 0.040 & 0.22 / 3.00 & 0.21 / 29.5 & 30.2 \\
\hline
\end{tabular}
\label{Tab:StarChar}
\begin{list}{}{}		         	                	                 
\item[$^{\mathrm{1}}$] Derived from spectra \hspace{0.5cm} $^{\mathrm{2}}$ Derived from Str\"omgren photometry
\end{list}{}{}		         	                	                 
\end{table*}

\section{Astrophysical parameters, chromospheric activity, and lithium abundance}
\label{Sec:Res}

High-resolution spectroscopic observations allow us to derive radial ($RV$) and projected rotational velocities ($v\sin i$), spectral type, luminosity class, and metallicity. They also enable us to identify spectroscopic binaries. Cross-correlation functions (CCFs) were computed to derive radial velocities (Table~\ref{Tab:SED_kin}) and we used the ROTFIT code \citep{2003A&A...405..149F,2006A&A...454..301F} to evaluate effective the temperature ($T_{\rm eff}$), gravity ($\log g$), metallicity ([Fe/H]), and $v\sin i$ of our targets (Table~\ref{Tab:StarChar}). We were also able to infer both the level of chromospheric activity (from the emission in the H$\alpha$ line core) and the age (from the lithium abundance). The equivalent width of the lithium line ($W_{\rm Li}$) and the net equivalent width of the H$\alpha$ line ($W_{\rm H\alpha}^{em}$) were measured in the spectrum obtained after subtracting the non-active template by integrating the residual H$\alpha$ and lithium profile. We refer to \cite{2006A&A...454..301F} and \cite{2009A&A...504..829G} for a more detailed presentation of the methods we used to derive lithium abundances, along with chromospheric and coronal luminosities (Table~\ref{Tab:StarChar}).
 
As an independent estimate of the main astrophysical parameters, we evaluated the effective temperature, radius, and metallicity of these stars from our Str\"omgren photometry using the $uvby\beta$ algorithm \citep{1985Ap&SS.117..261M}, which also provides an estimate of the color excess $E(b-y)$. As seen in Table~\ref{Tab:StarChar}, the values of effective temperature derived from spectroscopy and Str\"omgren photometry agree with each other to within 200\,K. The metallicity index $\delta m_0$ is in substantial agreement with the spectroscopic value [Fe/H]$\approx$\,0 for all stars. Estimated radii and $\log g$ values excluded the possibility of lithium-rich giant stars.

\section{Discussion}
\label{Sec:Disc}

\subsection{Star properties}

\begin{figure*}
\centering{\hspace{-0.6cm}\includegraphics[width=5.1cm]{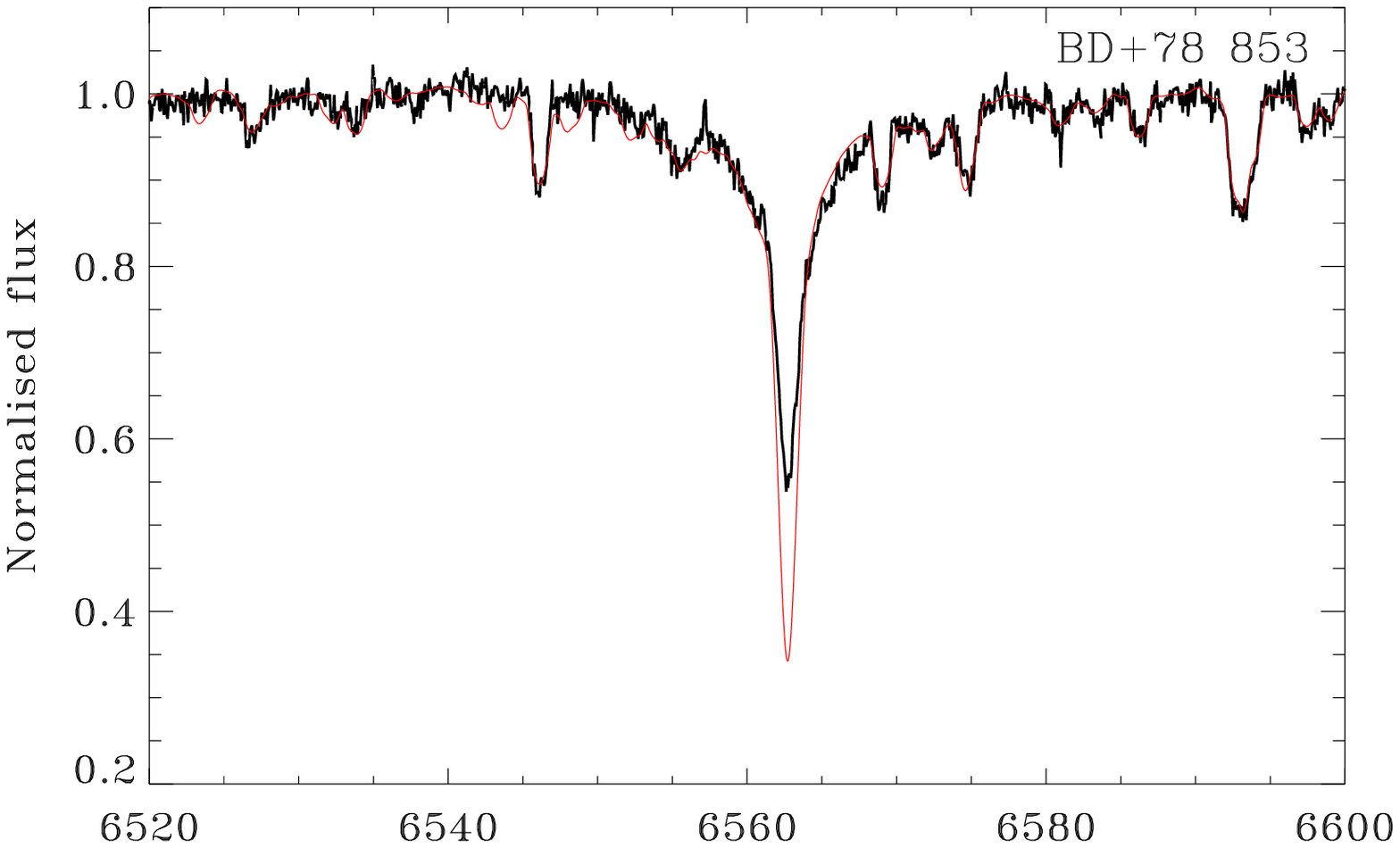}}  
\centering{\hspace{-0.6cm}\includegraphics[width=5.1cm]{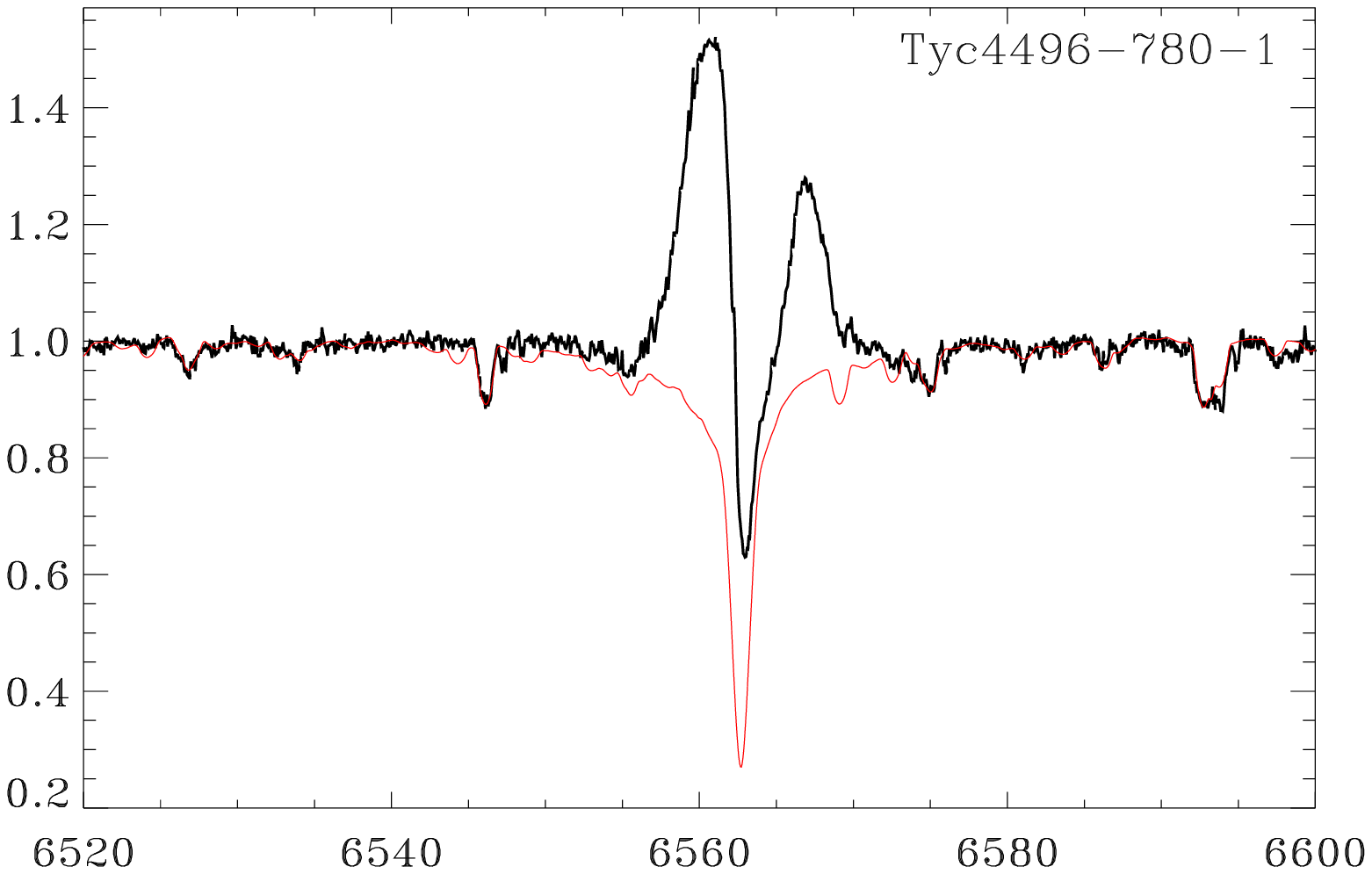}} 
\centering{\hspace{-0.6cm}\includegraphics[width=5.1cm]{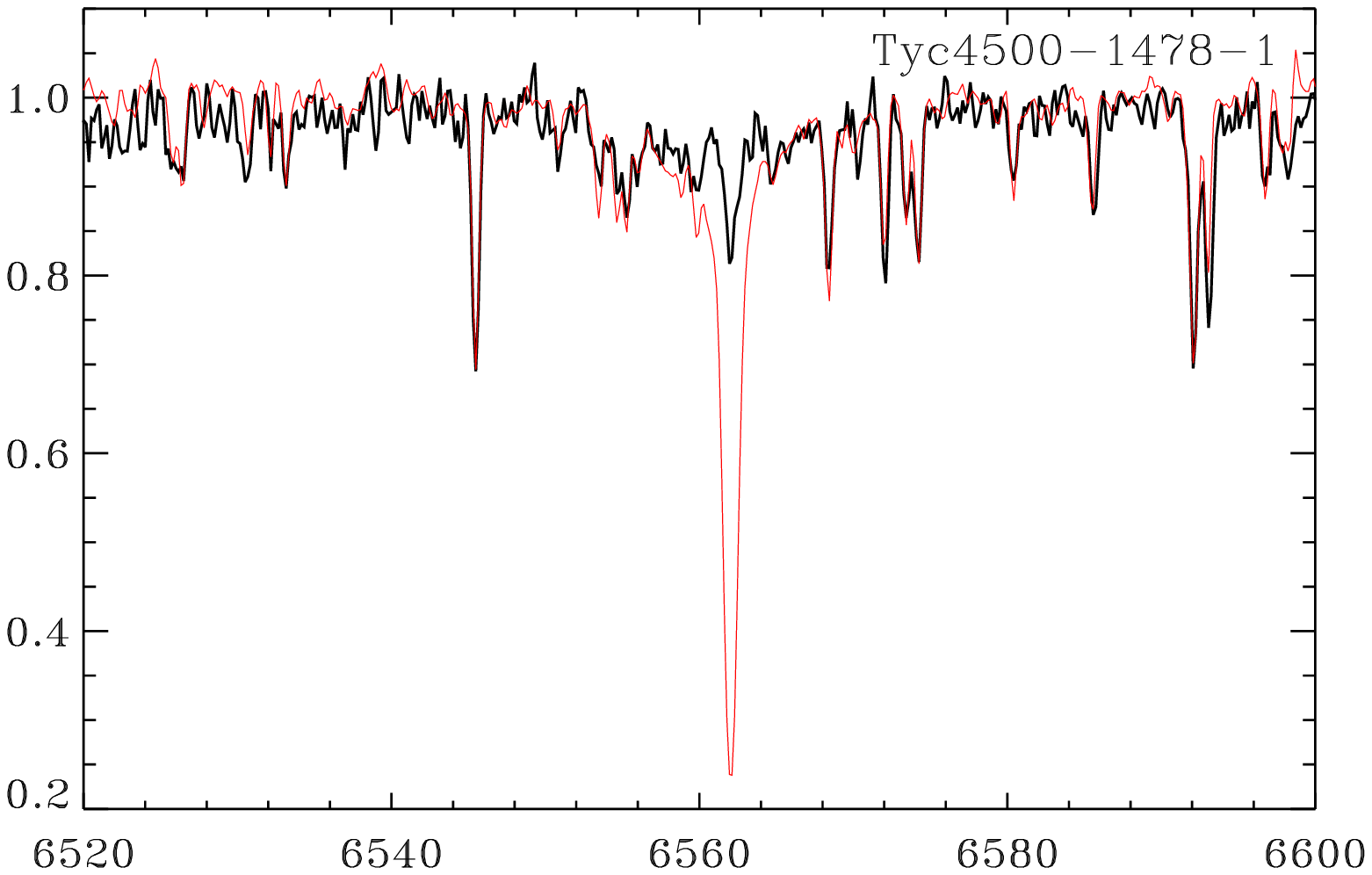}} 
\centering{\hspace{-0.6cm}\includegraphics[width=5.1cm]{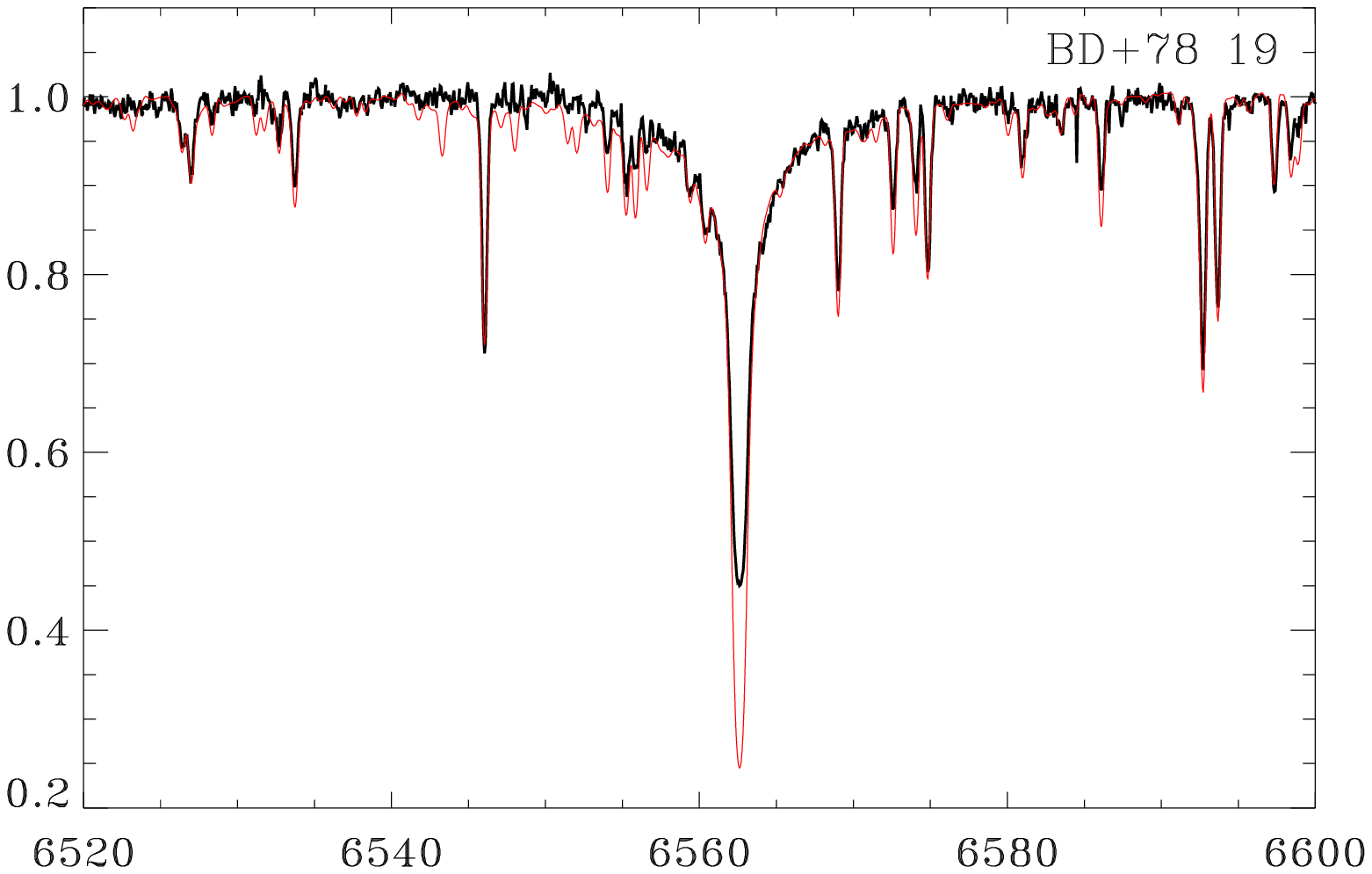}}\\   
\centering{\hspace{-0.6cm}\includegraphics[width=5.1cm]{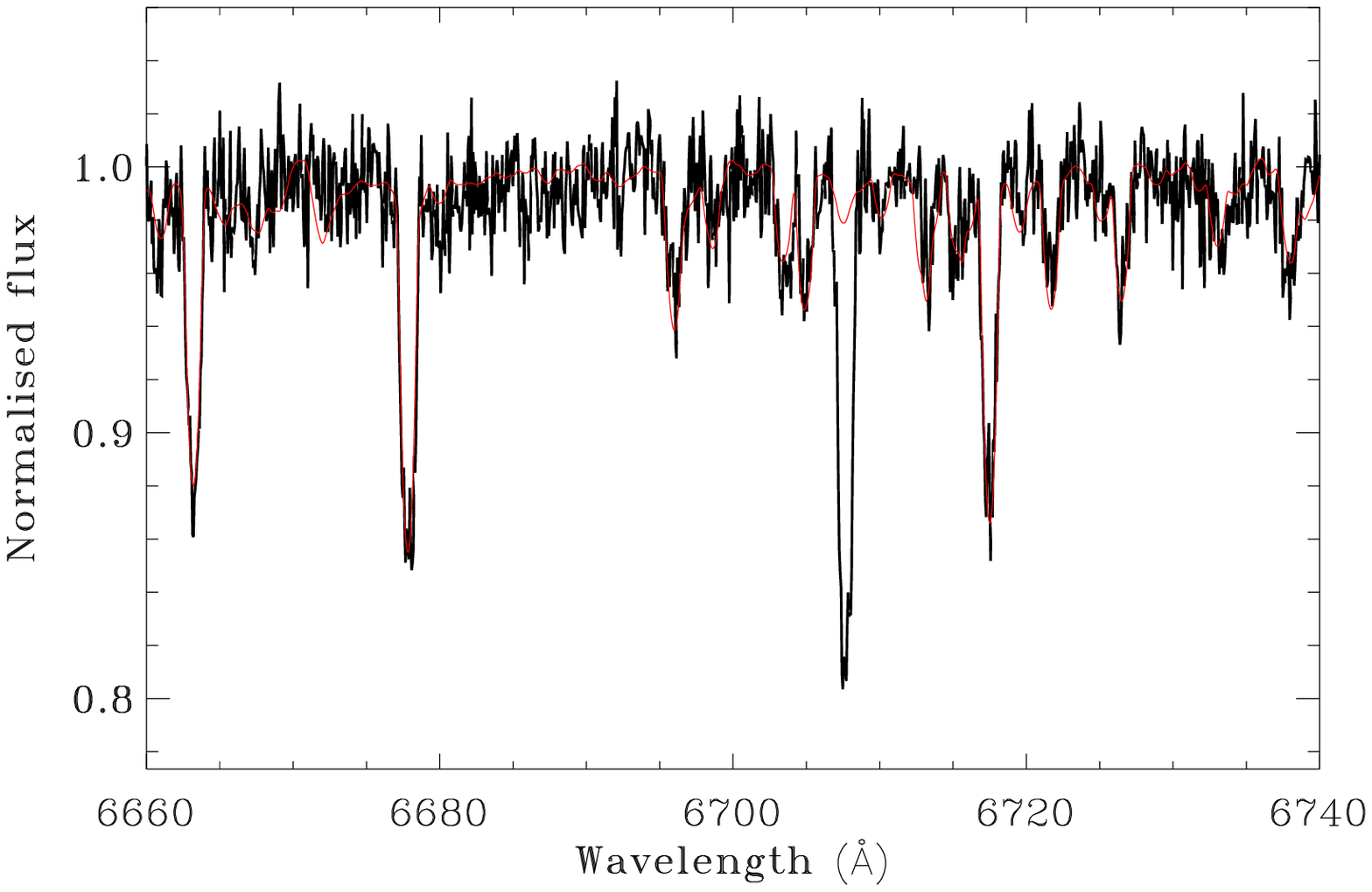}}  
\centering{\hspace{-0.6cm}\includegraphics[width=5.1cm]{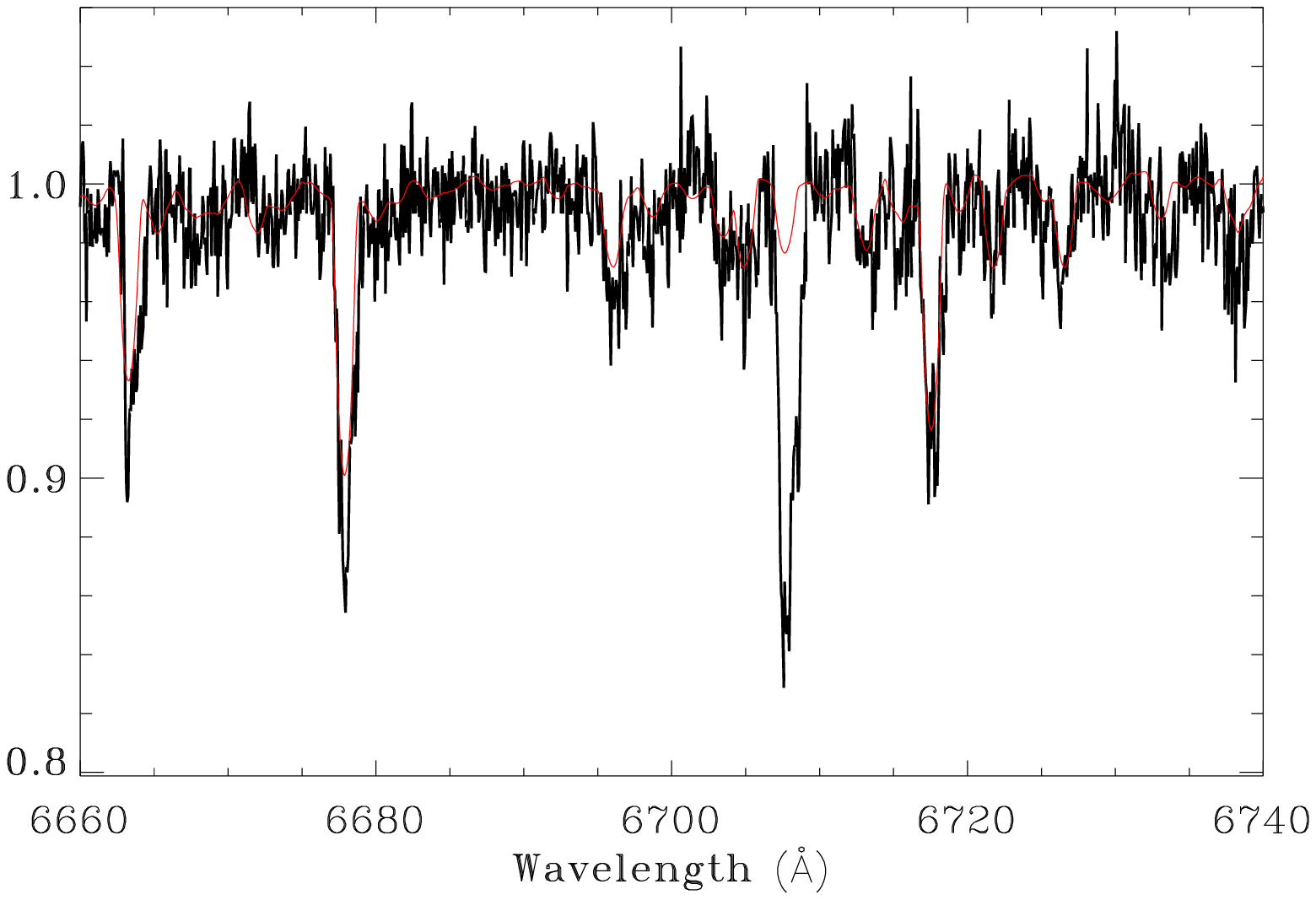}} 
\centering{\hspace{-0.6cm}\includegraphics[width=5.1cm]{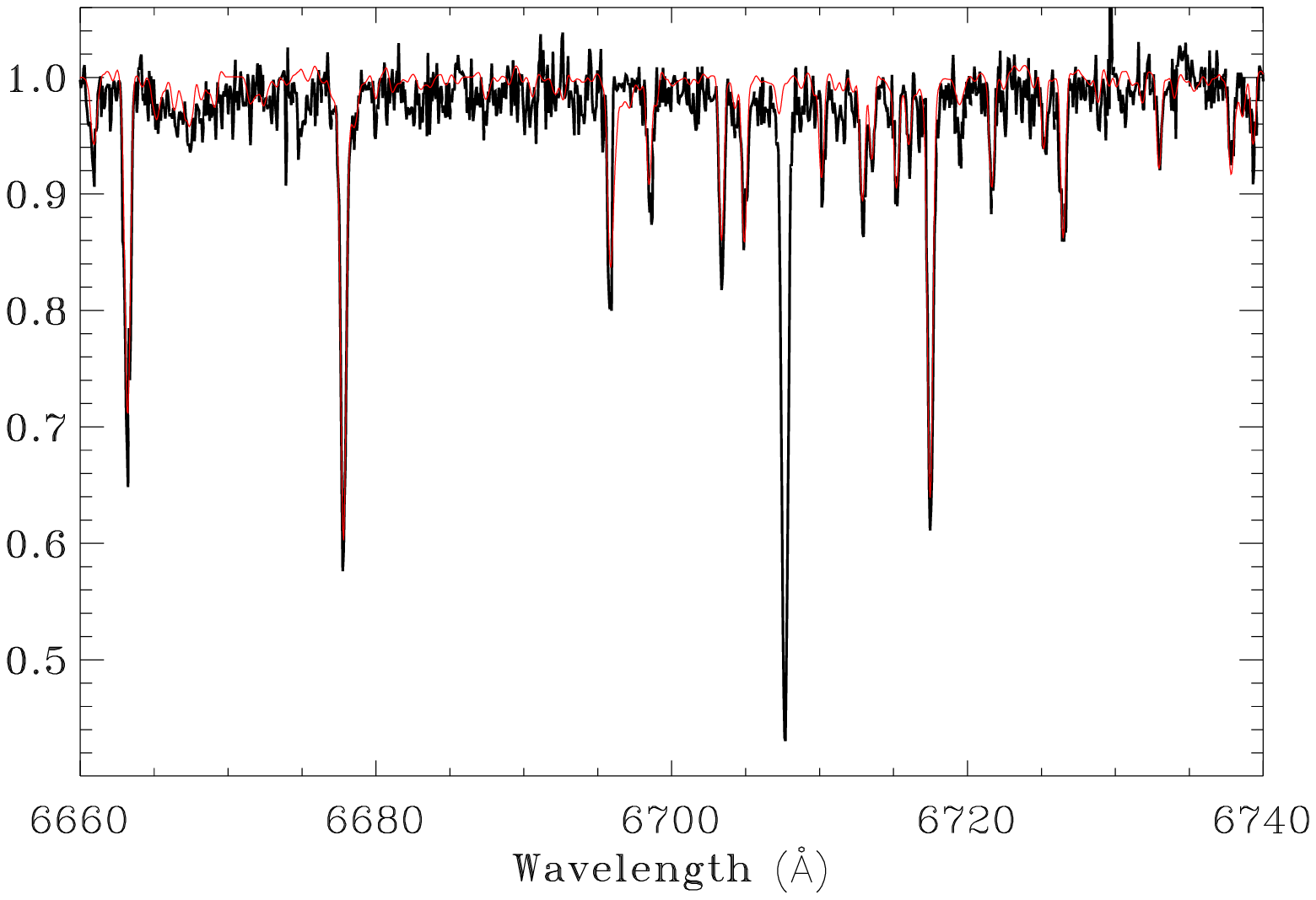}} 
\centering{\hspace{-0.6cm}\includegraphics[width=5.1cm]{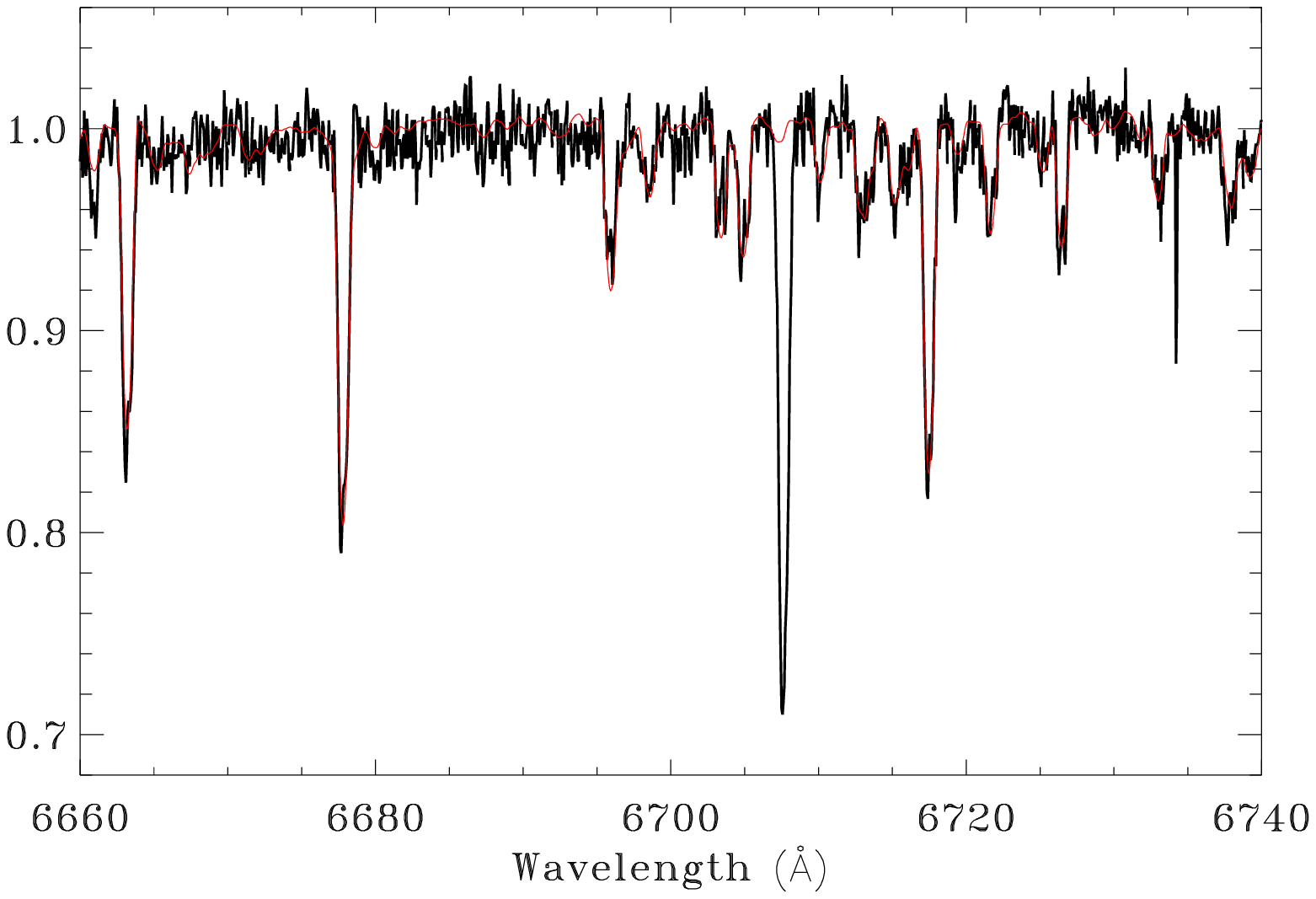}}	
\caption{Spectra (black line) in the H$\alpha$ (\textit{upper panels}) and lithium (\textit{lower panels}) region of the Cepheus comoving stars, and the template spectra (red lines) of non-active lithium-poor reference stars broadened to the $v\sin i$ of the targets and Doppler-shifted according to their $RV$.}
% ,bb=50 10 558 720,clip
  \label{Fig:PTTSspec}
\end{figure*}

The high resolution spectra and CCFs of our stars are shown in Figs.~\ref{Fig:PTTSspec} and ~\ref{Fig:PTTSCCF}\footnote{Available in electronic form only.} for both H$\alpha$ and \ion{Li}{i} spectral regions. The characteristics of the stars, derived from the analysis of the spectra acquired in the H$\alpha$ and lithium spectral regions, are now summarized below:\\

{\it BD+78\,853}: 
Although both the H$\alpha$ and lithium spectra display broad absorption lines, which are sometimes asymmetric, the automatic procedure failed to detect double peaks in any of the CCF. The H$\alpha$ profile is seen in absorption, but the analysis found that the line is partly filled. The equivalent width of the lithium line is $W_{\rm Li}$\,=\,0.18\,\AA, corresponding to a lithium abundance $\log N{\rm (Li)}$\,=\,3.05, which implies that BD+78~853 is young. The rotational velocity derived from the line broadening decreases by a factor two or so from the H$\alpha$ ($v\sin i\approx$~50\,km~s$^{-1}$) to lithium spectra ($v\sin i\approx$~28\,km~s$^{-1}$). This could be indicative of a spectroscopic binary, but may also be the result of a different CCF shape in the two spectral domains. The radial velocities derived from H$\alpha$ and \ion{Li}{i} spectral regions are compatible within 0.2 km~s$^{-1}$ at $RV=-6.4$\,km~s$^{-1}$. Our present data suggest that BD+78~853 is a single star that is rotating relatively quickly, but they do not exclude a spectroscopic binary system. More data are needed to settle this point.\\

{\it TYC 4496-780-1}:
The H$\alpha$ line exhibits an absorption feature roughly centered on a strong asymmetric H$\alpha$ emission profile. The \ion{Li}{i}\,$\lambda$6707.8 line with $W_{\rm Li}$\,=\,0.18\,\AA\ ($\log N{\rm (Li)}$\,=\,3.14) is stronger than the nearby calcium line, as in the case of BD+78~853. In both spectra, the lines are very broad and the code detected two Doppler-shifted dips (at $-5.5$ and 45.2\,km~s$^{-1}$) in the H$\alpha$ CCF with very different depths, suggesting a binary in which the two stars of the system have significantly different luminosities. In the lithium CCF, the footprint of a secondary component is clearly visible (although not detected by the automatic analysis) as an asymmetry on the red side of the main CCF dip, which is centered on about $-4$\,km~s$^{-1}$. We derived an $RV\approx$\,25\,km~s$^{-1}$ for the secondary dip, which would imply an $RV$ variation of $+2$ and $-20$\,km~s$^{-1}$ for the primary and secondary component, respectively, of this SB2 candidate. The projected rotational velocity of TYC~4496-780-1 is estimted to be $v\sin i$~=~$37~+/-10$\,km~s$^{-1}$.\\

{\it TYC 4500-1478-1}:
The FRESCO spectrum displays an H$\alpha$ profile markedly filled-in by emission (see Fig. 1). The lithium region shows very sharp absorption lines and exhibits a very strong \ion{Li}{i}\ $\lambda$6707.8 line with $W_{\rm Li}$\,=\,0.30\,\AA\ corresponding to $\log N{\rm (Li)}$\,=\,3.09. The perfect fit of the CCF allows us to derive accurate values of radial velocity ($RV=-8.8$\,km\,s$^{-1}$) and projected rotational velocity ($v\sin i \approx$~9\,km\,s$^{-1}$). Unless we have observed a binary system at a conjunction, we consider TYC 4500-1478-1 as a single star.\\

{\it BD+78\,19}:
The H$\alpha$ spectrum is typical of a mid-G type star without any remarkable characteristics. Both $RV=-9.7$\,km~s$^{-1}$ and $v\sin i \approx$~12\,km~s$^{-1}$ were derived from the Gaussian fit of the CCF. Apart from the observed \ion{Li}{i}\,$\lambda$6707.8 deep absorption line ($W_{\rm Li}$\,=\,0.22~$\AA$; $\log N{\rm (Li)}$~=~3.00), the lithium spectrum CCF is clearly not Gaussian and a fit of a rotational profile provides a closer match. The radial velocity determination from the lithium region is compatible with the former value but the rotational velocity is larger by a factor 2 or so. This suggests that, although the individual spectra appear as single-lined, BD+78\,19 may be a binary system of mass ratio $\approx$~1 observed with {\it Aurelie} close to conjunction at both epochs.\\

The chromospheric and coronal luminosities (see Table~\ref{Tab:StarChar}) were computed assuming that all stars are 15\,Myr old (see Table~\ref{Tab:SED_kin} for the adopted distances $d_{\rm 15Myr}$ and Sect.~\ref{Sec:Origin} for its justification). We found X-ray luminosities $L_X\,\simeq\,10^{30.4}$\,erg\,s$^{-1}$ for all sources (within 0.2 dex), which is typical of weak-line T Tauri stars (WTTS) in Taurus-Auriga-Perseus SFRs. For both BD+78\,853 and BD+78\,19, we found L$_{H\alpha}$\,=\,10$^{29.5}$\,erg\,s$^{-1}$, a value slightly higher than that found by \cite{2009A&A...504..829G} for their PTTS candidates and about an order of magnitude lower than that in old binary systems, in which the coupling of spin and orbital motions by tidal actions can maintain high chromospheric activity for a very long time \citep{2006A&A...454..301F}. The H$\alpha$ luminosity of TYC\,4496-780-1 is an order of magnitude higher than that of BD+78\,853 and BD+78\,19, but both the shape of the H$\alpha$ profile and the IR excess (see Sect.~\ref{Sec:SED}) infer that accretion is the primary cause of the observed emission.  
%----------------------------------------------------------- 
%----------------------------------------------------------- 
\begin{figure*}
\centering{\hspace{-0.6cm}\includegraphics[width=5.2cm,bb=54 379 558 698,clip]{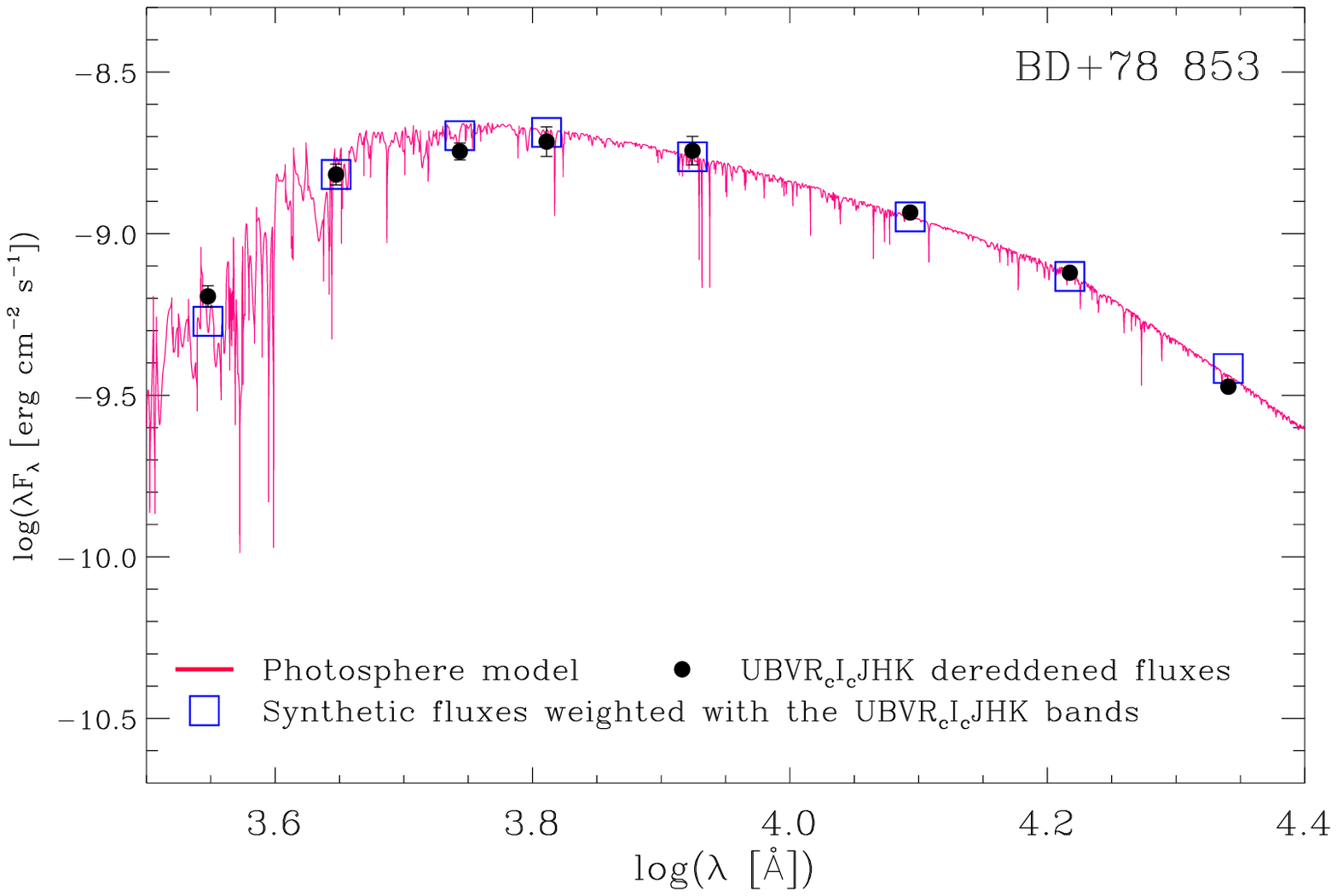}}  
\centering{\hspace{-0.2cm}\includegraphics[width=4.7cm,bb=110 379 558 698,clip]{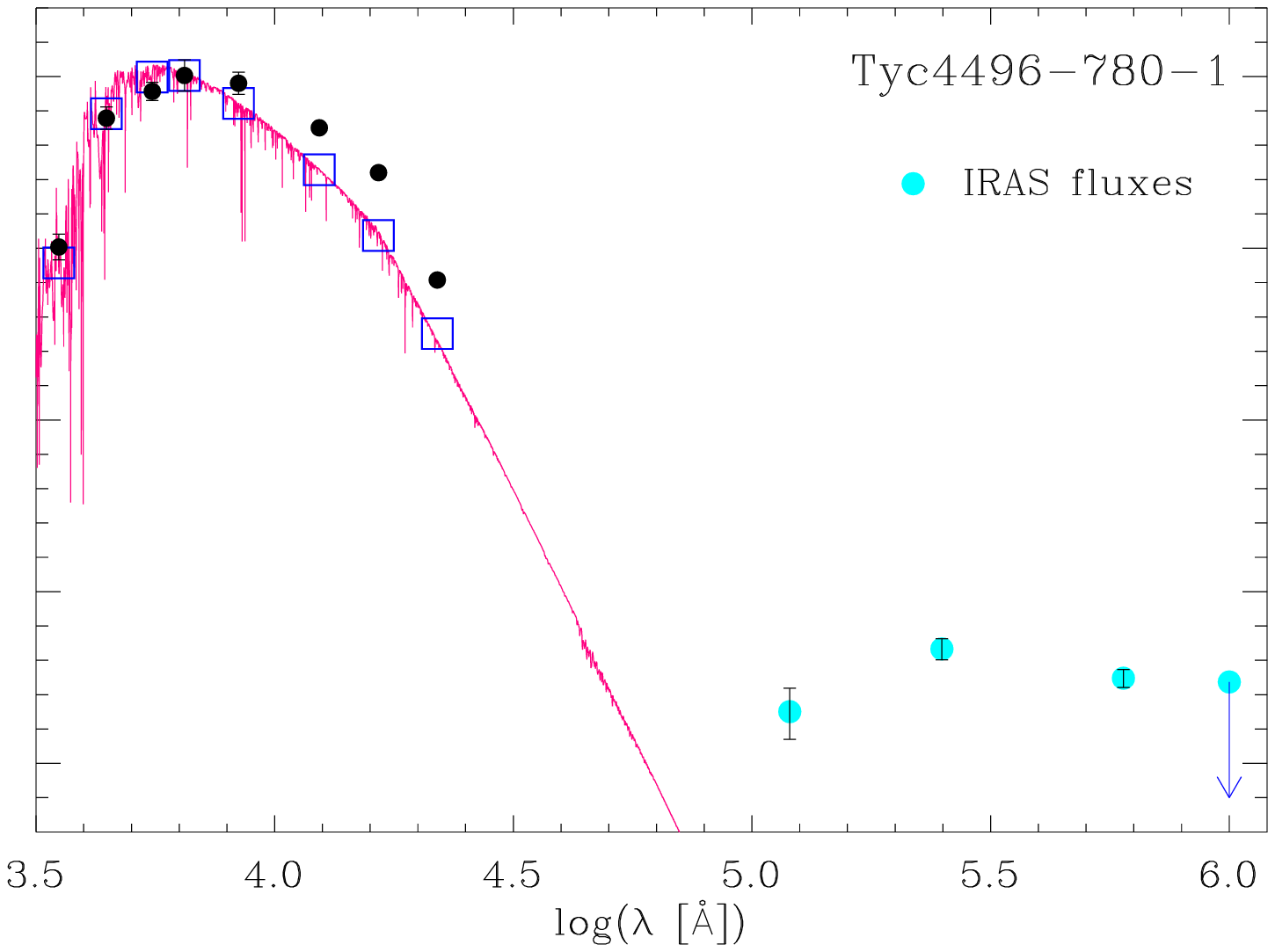}} 
\centering{\hspace{-0.2cm}\includegraphics[width=4.7cm,bb=110 379 558 698,clip]{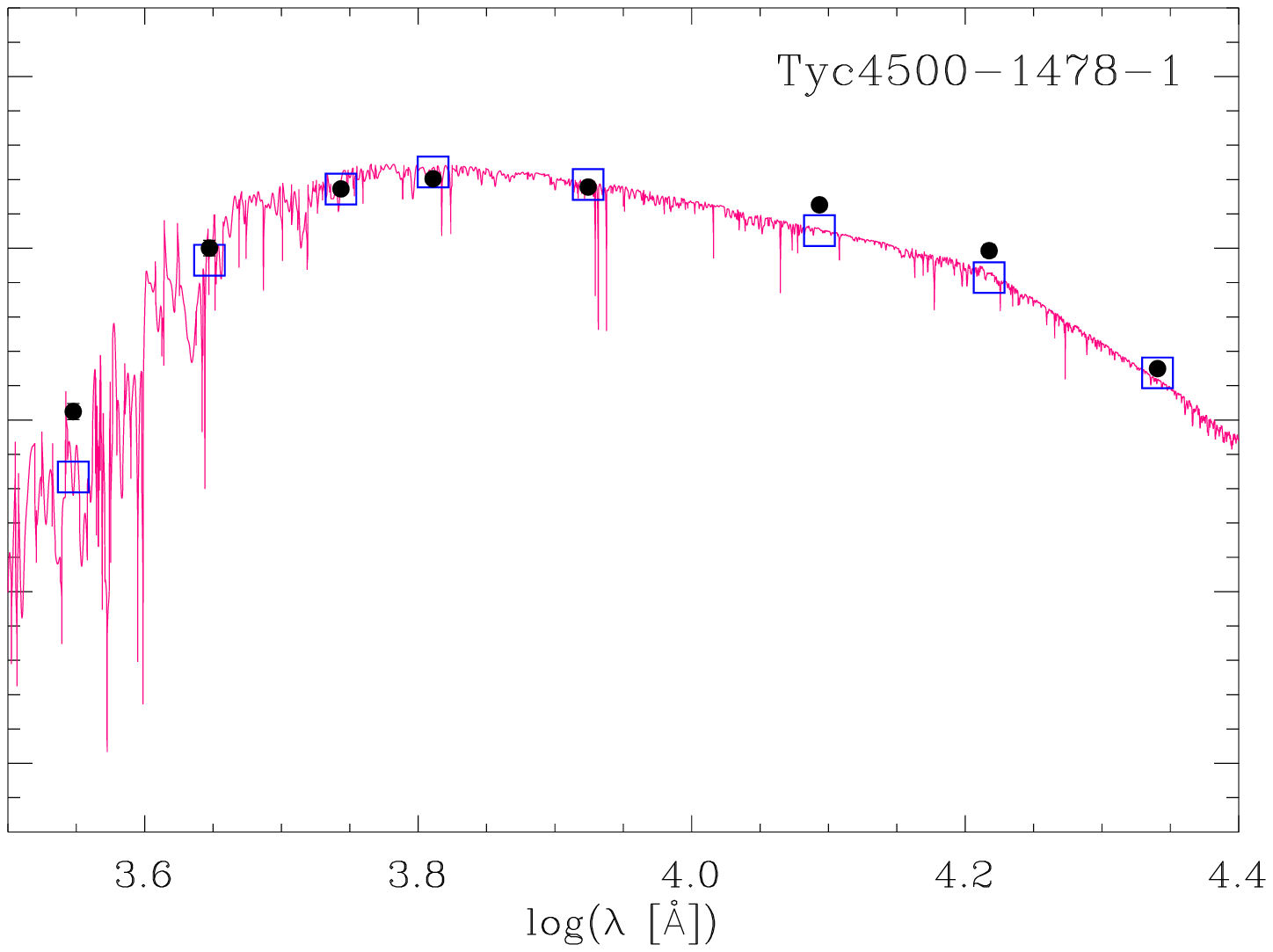}} 
\centering{\hspace{-0.2cm}\includegraphics[width=4.7cm,bb=110 379 558 698,clip]{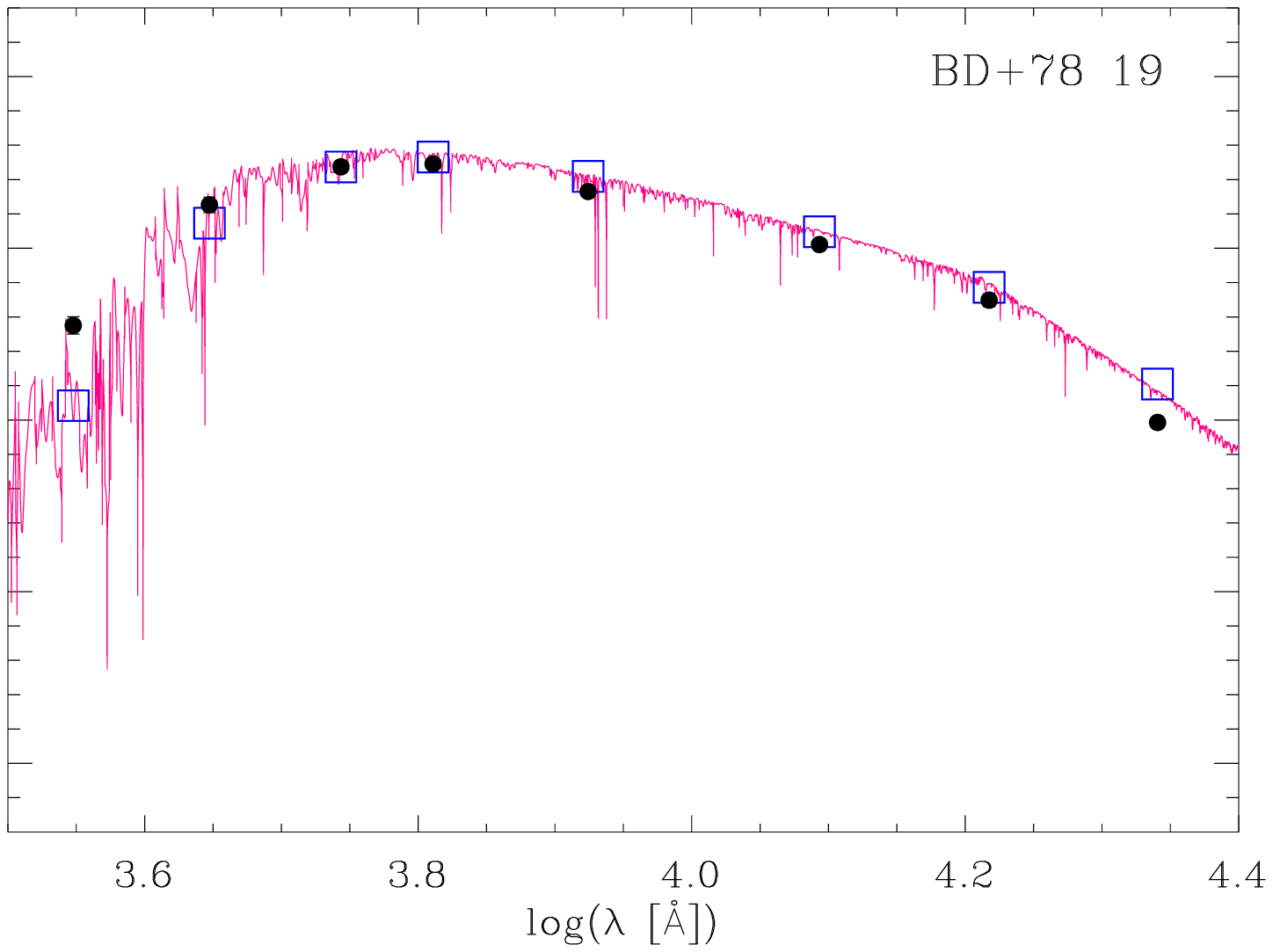}}   
\caption{Spectral energy distributions of the four stars (dots). The best-fit NextGen spectrum is overplotted with continuous lines in each box. } 
% ,bb=50 10 558 720,clip
  \label{Fig:SED}
\end{figure*}
%-----------------------------------------------------------
%
\subsection{Spectral energy distributions}
\label{Sec:SED}
\begin{table*}[t]
\caption{Parameters deduced from the SEDs and kinematics. The proper motions are from the TYCHO-2 catalogue.}
\begin{tabular}{lccccrcrrrrrrr}
\hline
\hline
Name         &  $\phi$ & $d_{\rm ZAMS}$ & $d_{\rm 15Myr}$ & $\mu_{\alpha}\cos\delta$ & $\mu_{\delta}$ & $RV$ & $U_{\rm ZAMS}$ & $V_{\rm ZAMS}$ & $W_{\rm ZAMS}$  & $U_{\rm 15Myr}$ & $V_{\rm 15Myr}$ & $W_{\rm 15Myr}$ \\
                &  (mas)  & (pc)	   & (pc)    & \multicolumn{2}{c}{(mas\,yr$^{-1}$)} & (km\,s$^{-1}$)   & \multicolumn{3}{c}{(km\,s$^{-1}$)}    & \multicolumn{3}{c}{(km\,s$^{-1}$)} \\ 
\hline
BD+78\,853       &  0.088  & 105 & 175 & $23.5$ & $-0.4$ & $-6.5$ & $7.0$ & $-10.5$ & $-4.0$ & 13.9 & $-13.9$ & $-6.0$ \\
TYC 4496-780-1  &  0.104  & 100 & 150 & $22.4$ &  $2.2$ & $-5.5$ & $6.5$ &  $-9.9$ & $-1.9$ & 11.1 & $-12.6$ & $-2.2$ \\
TYC 4500-1478-1 &  0.100  &  75 & 130 & $25.1$ & $-3.4$ & $-8.8$ & $3.2$ & $-11.5$ & $-4.1$ &  8.9 & $-14.6$ & $-5.3$ \\
BD+78\,19        &  0.092  &  90 & 155 & $23.2$ & $-2.8$ & $-9.7$ & $3.6$ & $-12.7$ & $-4.3$ &  9.8 & $-16.3$ & $-5.5$ \\
\hline
\end{tabular}
\label{Tab:SED_kin}
\end{table*}

The standard $UBVR_{\rm C}I_{\rm C}$ photometry (Table~\ref{Tab:StarPhot}), complemented with $JHK_{s}$ magnitudes from the 2MASS catalogue \citep{2003tmc..book.....C}, allowed us to reconstruct the spectral energy distribution (SED) from the optical to the near infrared (IR) domain for all sources. 

We used the grid of NextGen low-resolution synthetic spectra, with $\log g = 4.0$ and solar metallicity by \citet{1999ApJ...512..377H}, to perform a fit to the SEDs. The effective temperature ($T_{\rm eff}$) was kept fixed to the value derived with ROTFIT (Table~\ref{Tab:StarChar}) and we allowed the interstellar extinction ($A_V$), which affects the SED shape particularly at the shortest wavelengths, and the angular diameter, which scales the model surface flux to the stellar flux at Earth, free to vary. The \citet{1989ApJ...345..245C} extinction law with $R_V=3.1$ was used. The optimal solution was found by minimizing the chi-square of the fit. The latter was performed only on $UBVRIJ$ data, which are dominated by the photospheric flux of the star and are normally not appreciably affected by infrared excesses. The $A_V$ values are in the range 0.06--0.2\,mag, in agreement with the color excesses $E(b-y)$, which were inferred from the Str\"omgren photometry. The rather low $A_V$ values are typical of sources at $\approx$\,80--200\,pc that are not embedded in dusty nebulae, and consistent with their position far from dark clouds. The angular diameters ($\phi$), all on the order of 0.1 mas, are reported in Table~\ref{Tab:SED_kin}.\\

Figure~\ref{Fig:SED} displays the results of our fitting procedure. As is clearly evident, the SEDs are reproduced well by the synthetic spectrum blueward of the $K_{s}$ band, with the exception of TYC\,4496-780-1, which displays a strong near- and far-IR excess (IRAS fluxes). The lack of significant IR excess in three of our targets implies that these stars are likely to be either WTTS or PTTS. The slope of the SED between $K_{s}$ and 24\,$\mu$m, $\alpha \simeq -1.1$, allows us to tentatively classify TYC\,4496-780-1 as a Class\,II young infrared source, i.e. a T Tauri star surrounded by an accretion disc, according to the definition of \citet{1987IAUS..115....1L}. This classification is consistent with its H$\alpha$ emission profile. Similar H$\alpha$ line profiles are often observed in accreting TTS \citep{2005A&A...434.1005A}. According to the concept of the magnetospheric accretion model, the large line width (several hundred km~s$^{-1}$) is indicative of large-scale gas flows, while the inverse P Cygni profile traces the gas infall \citep{2001ApJ...550..944M,2006MNRAS.370..580K}.

\subsection{Kinematics and relation to Cep-Cas complex.}
\label{Sec:Origin}
\begin{figure}[htp]
\hspace{-0.7cm}
\includegraphics[width=9.6cm,bb=0 10 504 340,clip]{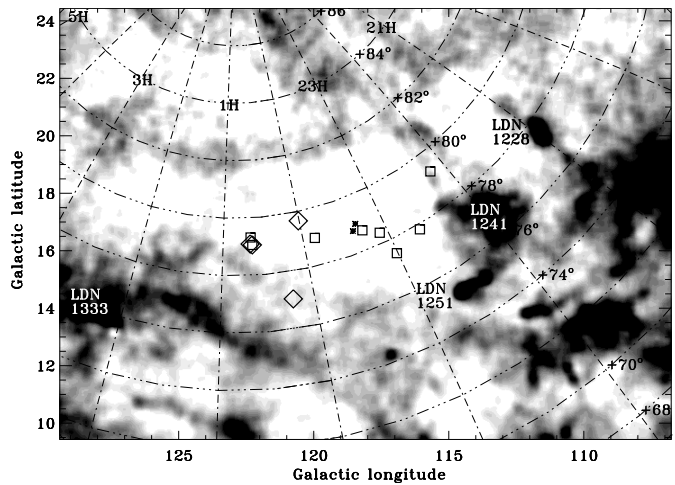}
\caption{Spatial distribution of our ``isolated'' comoving TTS (big diamonds) overplotted on a map of dark clouds enhancing interstellar material. The ``isolated'' young stars previously discovered in the CO void region are shown with square symbols, while the young ``naked'' PTTS HIP~115147 and its comoving companion are indicated by small asterisks. The major clouds in the vicinity are also labelled.}
\label{Fig:CoMVMap}
\end{figure}
The spatial distribution of BD+78\,853, TYC 4496-780-1, TYC 4500-1478-1 and, BD+78\,19 is shown in Fig.~\ref{Fig:CoMVMap} overplotted on the extinction map (Av) published by \citet{2005PASJ...57S...1D}, in a similar way to Fig.~2 of \cite{2005A&A...437..919T}. The position of the nearby young visual binary HIP\,115147 and its comoving companion, discovered by \cite{2007ApJ...668L.155M}, is also plotted together with the ``isolated'' young stars discovered by \cite{2005A&A...437..919T}. The closest prominent star-forming region is the Cep-Cas complex, which surrounds our stars. They are located in particular, at Galactic latitudes $14^\circ\,<\,b\,<\,17^\circ$, on the west side of the {\it Cepheus flare}, a region of significant extinction outside the Galactic plane first recognized by \cite{1934ApJ....79....8H}. Across the {\it Cepheus flare} region, \cite{1998ApJS..115...59K} determined the distances to several dark clouds and concluded that the interstellar matter is concentrated at 200, 300, and 450\,pc (although the whole Cep-Cas complex extends to 800\,pc or so) separated roughly at regular intervals along Galactic latitudes. \cite{2006MNRAS.369..867O} suggested that the {\it Cepheus flare} forms part of a large shell, 50\,pc in radius, expanding at 4~km~s$^{-1}$, that encloses an old supernovae remnant \citep{1989ApJ...347..231G}, whose center can be located at $l\,=\,120^\circ, b\,=\,17^\circ$. The {\it Cepheus flare} is a region of active star formation as demonstrated by \cite{2009ApJS..185..451K}, who discovered about 70 new PMS stars in the star-forming clouds.

All the four stars investigated by ourselves are projected in front of the void located between the dark clouds LDN~1333 (west), LDN~1241-1251 (east), and north of LDN~1259-1262. A group of seven stars, including TYC\,4500-1478-1, located in this void and displaying typical T Tauri characteristics, was identified by \cite{2005A&A...437..919T}. For a few stars, including TYC\,4500-1478-1, they found a lithium equivalent width close to that often measured at the Pleiades upper envelope, implying a PTTS or ZAMS evolutionary status for them. Although they consider these stars to be associated with the Cep-Cas complex, only on the basis of their proximity to this region, and adopt a distance of 200\,pc for them, their relation to the cloud complex remains uncertain.\\

The distance of our stars is crucial in determining their origin and a possible association with the Cep-Cas complex. Unfortunately, the Tycho parallaxes are useless and one must rely on photometric distance estimates. To cover all the possibilities, we estimated for each star two distances that we assume to be a lower and upper limit. The lower limit was calculated for the hypothesis that the star is on the ZAMS. The star radius of a ZAMS star with the same $T_{\rm eff}$ was adopted and the distance was deduced from the angular diameter $\phi$ reported in Table~\ref{Tab:SED_kin}. The upper limit was estimated assuming a minimum age and the corresponding radius was computed from the \citet{2000A&A...358..593S} evolutionary tracks. An age of $5$\,Myr would place stars at distances mostly exceeding 250\,pc, i.e. farther than the closest boundary of the {\it Cepheus flare} shell in that direction. This would contradict our photometric observations that no stars experience significant interstellar extinction. An age older than $\approx$\,20-30\,Myr would contradict the observational finding that, at this age, stars display no significant NIR excess \citep{2006AAS...209.1001H,2006PASP..118.1690M}. A good compromise is $15$\,Myr that we adopted in the following. The lower and upper limits to the distances ($d_{\rm ZAMS}$ and $d_{\rm 15Myr}$, respectively) are reported in Table~\ref{Tab:SED_kin}.

As seen in this table, the {\it ZAMS assumption} would place our stars roughly within 100\,pc of the Sun ruling out their association with the Cep-Cas complex whose CO clouds are farther away by nearly a factor of two. Needless to say that this assumption is also hardly compatible with the detection of the IR excess from an accretion disc around TYC 4496-780-1. On the other hand, the {\it 15\,Myr assumption} would place them roughly at the distance of LDN\,1333, LDN\,1261, and LDN\,1228 (see Fig.5 of \citealt{2008hsf1.book..136K}) and can thus partly reconcile their possible association with at least the nearest clouds of the {\it Cepheus Flare}.\\

\begin{figure}[t]
\includegraphics[width=7.0cm,bb=0 15 566 540,clip]{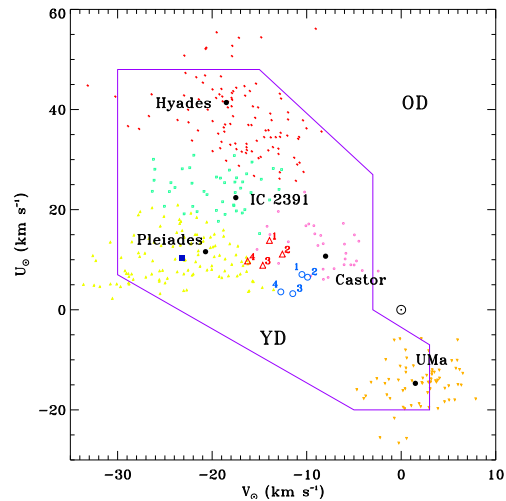}
\caption{U$-$V kinematic diagram for our comoving stars (labelled as in Table~\ref{Tab:MVCand}). Blue circles and red triangles refer to the {\it ZAMS} and {\it 15\, Myr assumptions}, respectively. The average velocity components (dots) of some young SKG and those of some late-type stars members of these young SKG are also plotted (square, triangle, circle, upside -down triangle, and U symbols for the IC~2391 supercluster, Pleiades, Castor, UMa moving groups, and Hyades supercluster, respectively). The loci of the young-disc (YD) and old-disc (OD) populations are also marked. The position of HIP\,115147 is plotted as a filled square.}
\label{Fig:CoMVkin}
\end{figure}

\begin{figure*}[htp]
\includegraphics[width=4.5cm]{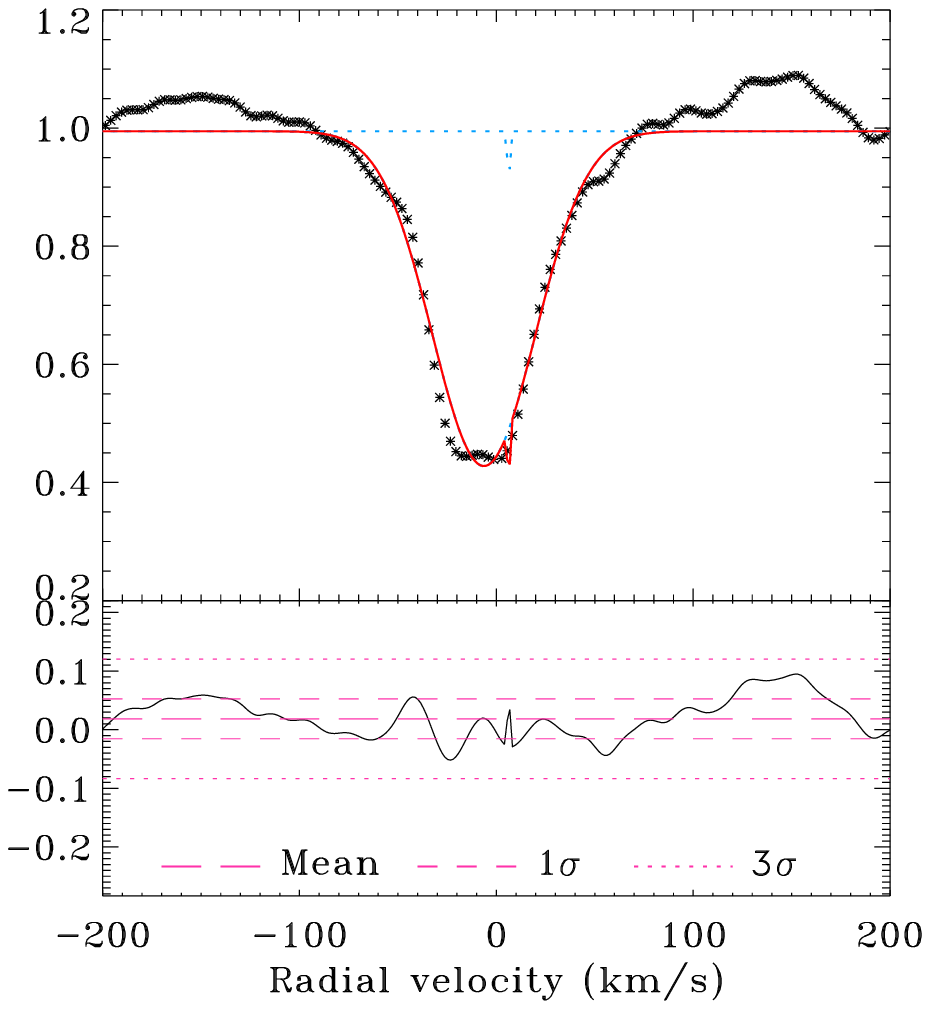}
\includegraphics[width=4.5cm]{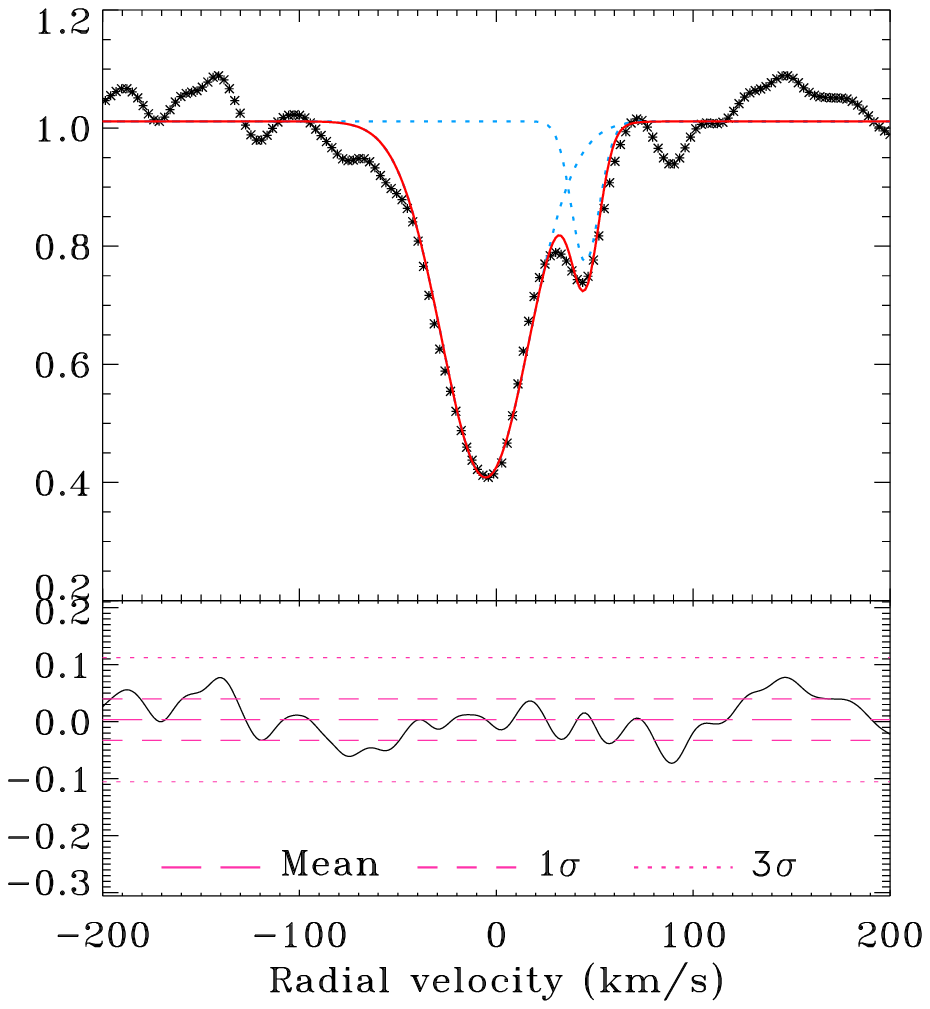}
\includegraphics[width=4.5cm]{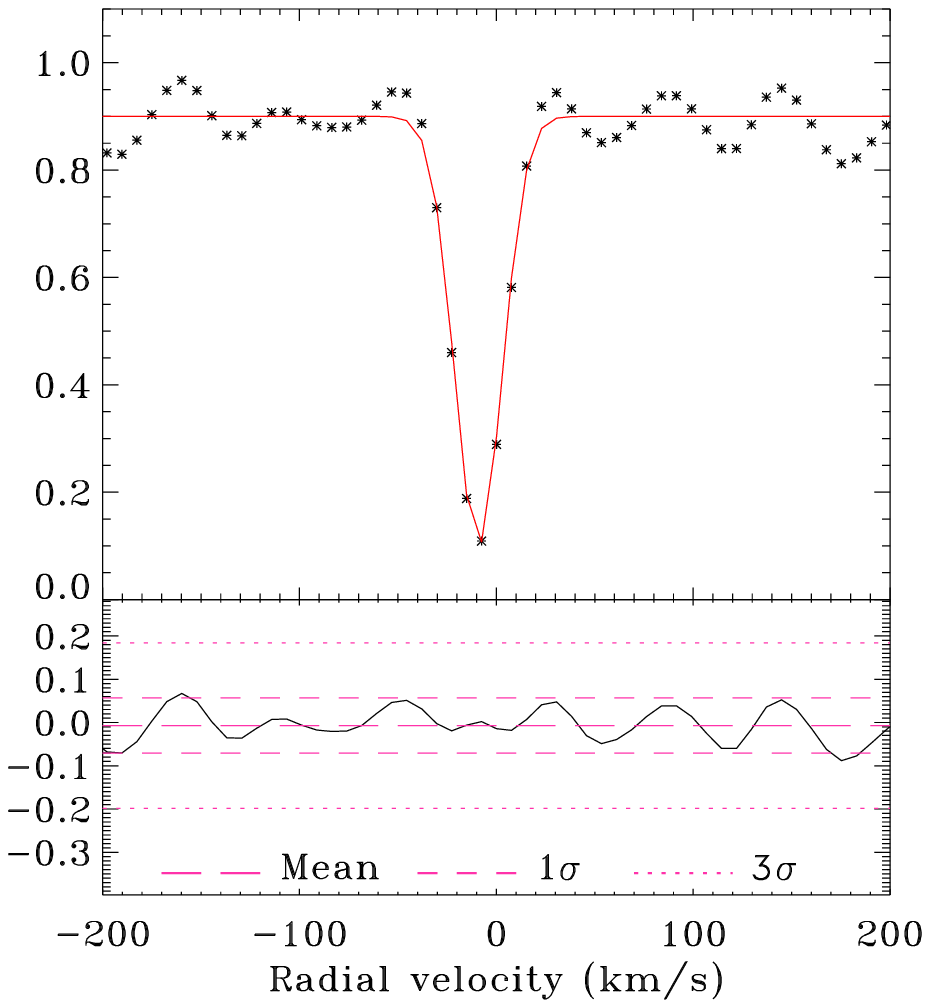}
\includegraphics[width=4.5cm]{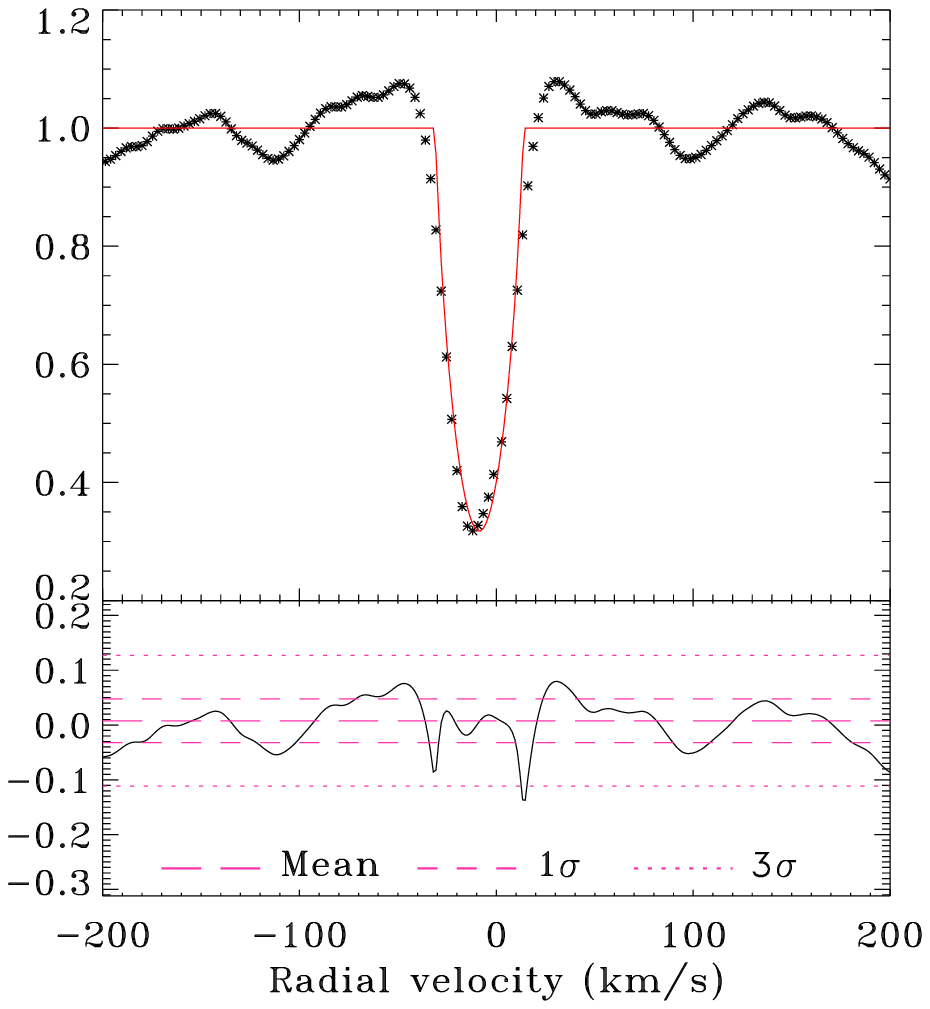}\\
\includegraphics[width=4.5cm]{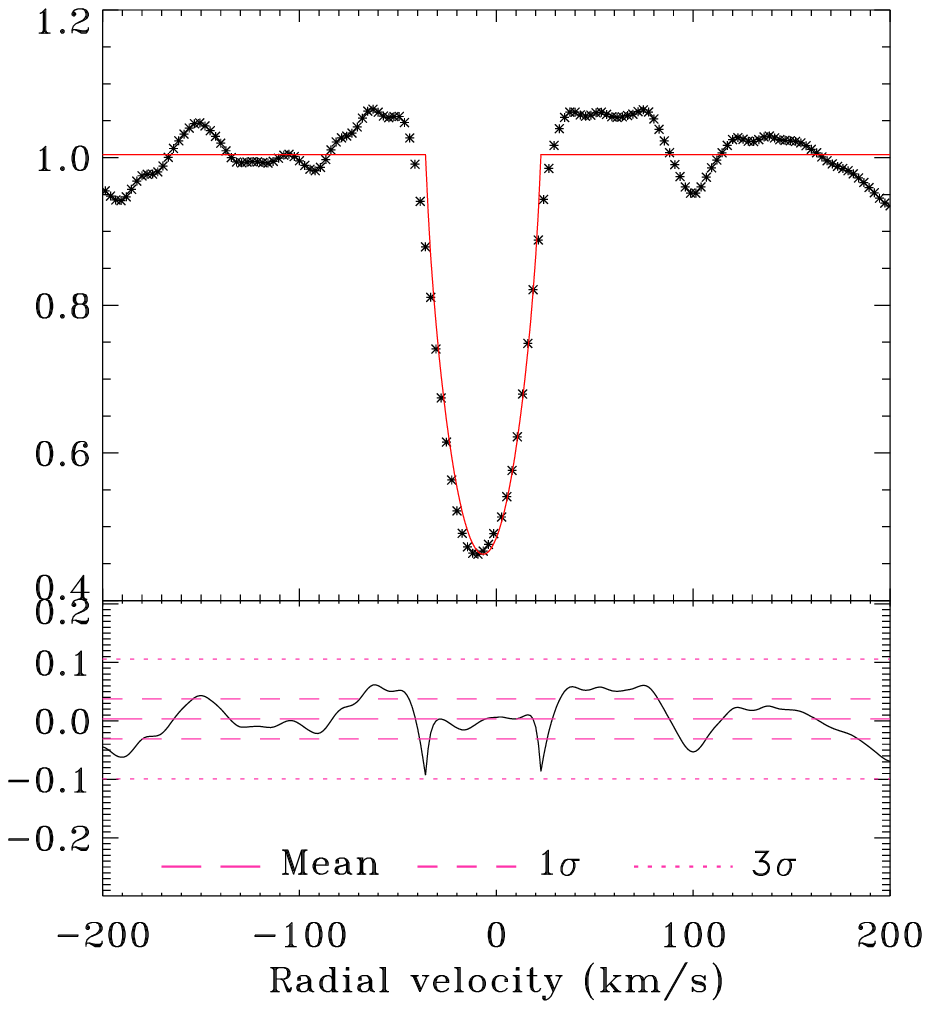}
\includegraphics[width=4.5cm]{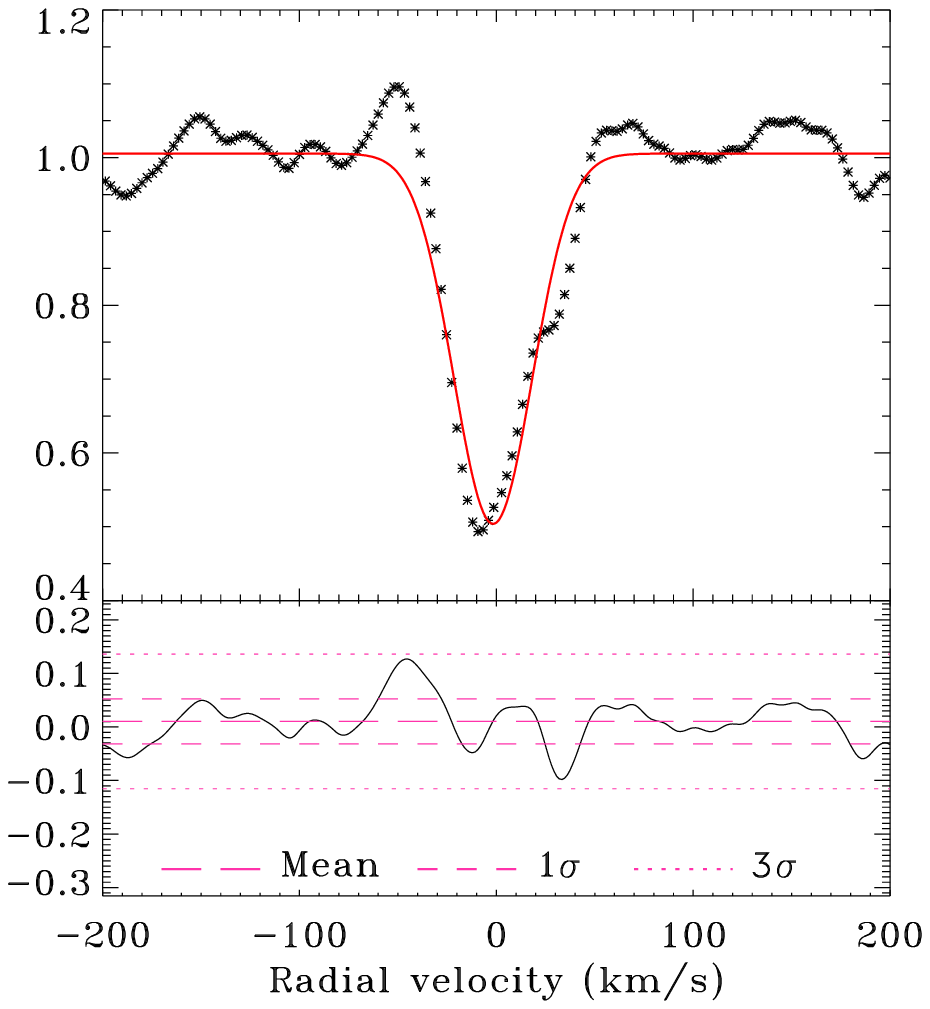}
\includegraphics[width=4.5cm]{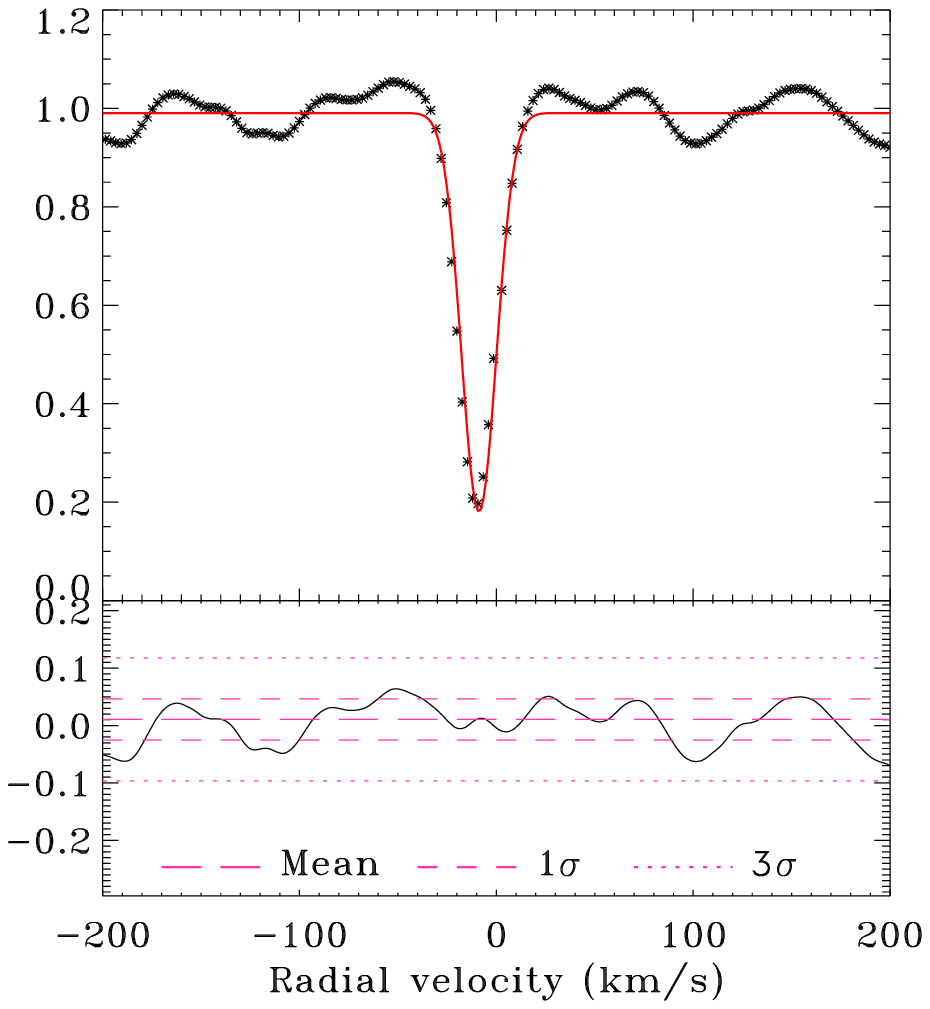}
\includegraphics[width=4.5cm]{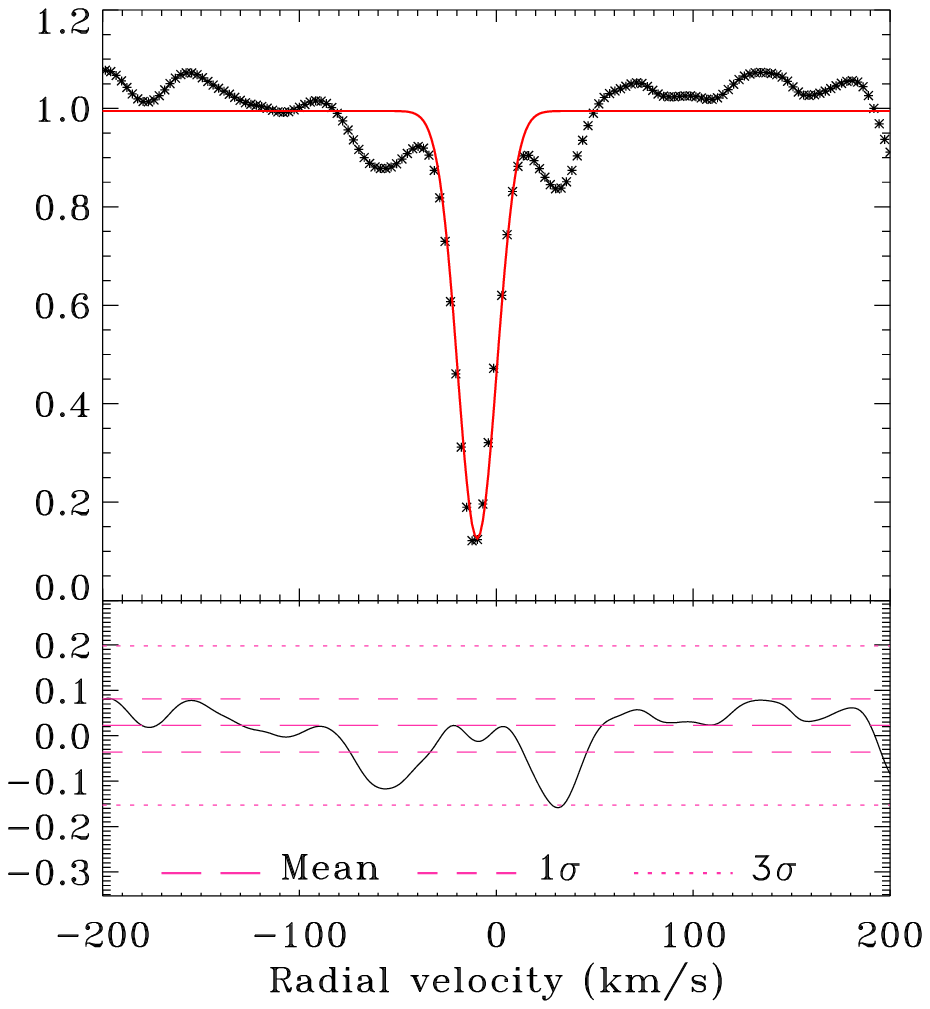}
\caption{Cross-correlation functions (cross symbols) of the Cepheus moving groups TTS candidates BD+78\,853 (RasTyc~0000+7940), TYC 4496-780-1 (RasTyc~0013+7702), TYC 4500-1478-1 (RasTyc~0038+7903), and BD+78\,19 (RasTyc~0039+7905) (from left to right) computed from their H$\alpha$ (\textit{upper panels}) and lithium (\textit{lower panels}) spectra. The continuous red line displayed in each panel shows the best fit, either with a Gaussian or a rotational profile. The residuals are displayed with a thin black line in the bottom of each panel.}
\label{Fig:PTTSCCF}
\end{figure*}

Additional crucial information to elucidate their origin tentatively can be derived from kinematics. Because errors in the distances are the major uncertainties (proper motions and radial velocity errors are smaller than or equal to 2~mas and 1.5~km~s$^{-1}$ respectively), we computed the heliocentric space velocity components, $UVW$, in a left-handed coordinate system, both with distances corresponding to ZAMS radii ($U_{\rm ZAMS}V_{\rm ZAMS}W_{\rm ZAMS}$) and adopting the $d_{\rm 15Myr}$ distances ($U_{\rm 15Myr}V_{\rm 15Myr}W_{\rm 15Myr}$), i.e. assuming that all stars are located in proximity to the closest clouds of the Cep-Cas complex. The $RV$ values measured on our spectra and Tycho-2 proper motions \citep{2000A&A...355L..27H} have been used. The positions of our four stars and HIP\,115147 in the $U$-$V$ kinematical diagram are shown in Fig.~\ref{Fig:CoMVkin}, together with the mean position of the major stellar kinematics groups (SKG) discussed in \citet{2001MNRAS.328...45M}, namely the IC~2391 supercluster ($\sim 50$~Myr), the Pleiades SKG ($\sim 100$~Myr), the Castor SKG ($\sim 200$\,Myr), the Ursa Major (UMa) group ($\sim 300$~Myr), and the Hyades supercluster ($\sim 600$\,Myr). Figure~\ref{Fig:CoMVkin} readily shows that, whatever the distance is, all our stars share the same kinematics (within a few km~s$^{-1}$) proving that they form a homogeneous MG with the same origin. We also note that both distances and kinematics seem to rule out an origin in common origin with HIP\,115147.\\
To assess their membership to the aforementioned SKGs on more objective grounds, we used the probabilistic approach described in \cite{2008PhDT.........7V}, i.e. we computed the probability $P$ that a star with heliocentric velocities $U$, $V$, and $W$ has a Galactic motion compatible with a given SKG. Independently of the distance assumption, the quantitative analysis excludes any possible association with the oldest UMa and Hyades MGs ($P\,\approx\,0$), as expected from the high-lithium photospheric abundance of our stars. Table~\ref{Tab:MVCand} suggests a possible link with the Castor MG, which is, however, again not supported by the very high-lithium abundance measurements, hardly compatible with a 200\,Myr old MG. With a less than 5\,\% probability, the IC~2391 MG origin is highly improbable. We are left with the Pleiades SKG for which the probabilities rise from $\approx$\,15\,\% for the {\it ZAMS assumption} to $\approx$\,40\,\% for the {\it 15\,Myr assumption}. However, as recalled by \cite{2007ApJ...668L.155M}, this stream includes isolated stars, groups, and associations of diverse ages in the range from 1\,Myr to about 200\,Myr, which by itself, does not shed light on the origin of our stars.\\
\begin{table}[t]
\caption{Membership probability $P$ (\%) of our four stars to major MGs computed with the {\it ZAMS} (left) and {\it 15\,Myr assumption} (right).}
\begin{tabular} {llccc}
\hline \hline
\# & Name & IC~2391 & Pleiades & Castor \\ 
\hline
1 & BD+78\,853 & 0.0 / 4.7 & 15.9 / 36.9 & 59.6 / 41.7\\
2 & TYC 4496-780-1 & 0.0 / 0.3  & 8.9 / 22.7 & 33.4 / 28.2 \\
3 & TYC 4500-1478-1 & 0.0 / 0.1 & 11.3 / 50.5 & 33.2 / 32.8 \\
4 & BD+78\,19 & 0.0 / 0.3 & 17.4 / 65.9 & 28.6 / 20.0 \\
\hline
\end{tabular}
\label{Tab:MVCand}
\end{table}

We finally consider a possible link with the Cep-Cas complex. Projected on the sky, the closest dark clouds to our stars are LDN\,1251 and LDN\,1241 at both 5--8$^\circ$ east of them and about 300\,pc from the Sun \citep{2008hsf1.book..136K}. The average proper motions of LDN\,1251, as estimated thanks to its cloud members discovered by \cite{2009ApJS..185..451K}, are $\mu_{\alpha}\,=\,1.9\,\pm\,1.7$ and $\mu_{\delta}\,=\,-0.5\,\pm\,1.3$\,mas\,yr$^{-1}$. The angular distance of 5--8$^\circ$ corresponds to 25--40\,pc at the cloud distance. With a proper motion of about 21 mas\,yr$^{-1}$ relative to LDN\,1251, nearly aligned in the east-west direction and corresponding to a velocity of about 30\,km\,s$^{-1}$ at the cloud distance, they should have travelled for 0.8--1.3\,Myr from their parental cloud. The closest cloud on the approaching side of the expanding shell (namely LDN\,1228, $d\simeq 180$\,pc, \citealt{2008hsf1.book..136K}) is about 10$^\circ$ east of our stars ($\approx$\,30\,pc at the cloud distance). By completing the same calculations for LDN\,1251/1241, we found an escape velocity ($v_{esc}$) from LDN\,1228 of about 22 km\,s$^{-1}$ and a travel time of about 1.3\,Myr. These estimates do not allow us to exclude that these stars are ``runaway'' objects that originated in LDN\,1251, LDN\,1241, or LDN\,1228. Nevertheless, although a close encounter with massive stars can result in high velocity ``runaway'' stars, \cite{1998A&A...339...95S} found that 60\,\% of the runaways have speeds larger than 3\,km\,s$^{-1}$ but that this fraction steeply decreases, becoming smaller than 5\,\% for $v_{esc}\,>$\,10\,km\,s$^{-1}$. Both the high escape velocites and the common space motions cast doubt on the ``runaway'' hypothesis being applicable.\\

The more plausible explanation of how these ``isolated'' TTS  formed is given by the ``in-situ'' model \citep{1996ApJ...468..306F}. This scenario would support the conclusion of \cite{2005A&A...437..919T} although the lack of accurate distances does not allow us to draw firm conclusions about the origin of this group of stars. The advent of GAIA mission, with its unprecedented astrometric precision, will certainly shed light on this open issue. Similar conclusions have been reached in the Taurus \citep{1997A&AS..124..449M}, Chamaeleon-Musca \citep{1998ApJ...507L..83M} and Lupus \citep{2001PASJ...53.1081T} SFRs although no accreting TTS have been found outside these SFR's cores.

\section{Conclusion}
\label{Sec:Conc}

We have reported the discovery of four comoving very young stars located in a region devoid of dark matter and molecular clouds. On the basis of its H$\alpha$ profile and infrared excess, one of the stars investigated has been tentatively classified as a Class\,II young infrared source, i.e. a TTS surrounded by an accretion disc. Owing to the distance uncertainty, we cannot assert that these stars are completely unrelated to the Cep-Cas star-forming region. However, their kinematics prove that they form a homogeneous comoving group of stars with the same origin. Their off-cloud positions imply that they are very good ``isolated'' TTS candidates. The ``runaway'' hypothesis is highly improbable because of their kinematical properties. Our conclusions raise a question about the applicability of the ``in-situ'' star-formation scenario to very low-mass cloud environments. TYC 4496-780-1 could be a TW Hya analog, although a little bit older, and its comoving companions may represent the peak of the iceberg of a young loose association. 

\begin{acknowledgements}
We are grateful to the OHP night assistant staff in conducting our Key Program, and those of the OAC observatories for their support and help with the observations. We thank Luigia Santagati of INAF-Catania for the English revision of the text. This research made use of SIMBAD and VIZIER databases, operated at the CDS, Strasbourg, France. This publication uses ROSAT data. Part of this work was supported by the Universidad Complutense de Madrid (UCM), the Spanish \emph{MICINN, Ministerio de Ciencia e Innovaci\'on} under grant AyA2008-00695 and the \emph{Comunidad Aut\'onoma de Madrid}, under PRICIT project S-2009/ESP-1496 (AstroMadrid).
\end{acknowledgements}

\bibliographystyle{aa}
\bibliography{RefBib}

\Online

\end{document}